
\input harvmac \noblackbox

\def\A{{\cal A}}
\def\B{{\cal B}}
\def\K{{\cal K}}

\def\d{\partial_{\pm}}


%
%
\def\RF#1#2{\if*#1\ref#1{#2.}\else#1\fi}
\def\NRF#1#2{\if*#1\nref#1{#2.}\fi}
\def\refdef#1#2#3{\def#1{*}\def#2{#3}}
\def\rdef#1#2#3#4#5{\refdef#1#2{#3, `#4', #5}}

%
%
\def\ts{\hskip .16667em\relax}

\def\CMP{{\it Comm.\ts Math.\ts Phys.\ts}}

\def\FAP{{\it Funct.\ts Analy.\ts Appl.\ts}}
\def\IJMP{{\it Int.\ts J.\ts Mod.\ts Phys.\ts}}

\def\JAP{{\it J. Appl.\ts Phys.\ts}}

\def\JP{{\it J.\ts Phys.\ts}}

\def\NP{{\it Nucl.\ts Phys.\ts}}
\def\PL{{\it Phys.\ts Lett.\ts}}

\def\Zm{Zamolodchikov}
\def\AZm{A.B. \Zm}

\def\ped{P.\ts E.\ts Dorey}
\def\dur{H.\ts W.\ts Braden, E.\ts Corrigan, \ped\ and R.\ts Sasaki}
%
%
\rdef\rAFZa\AFZa{A.E. Arinshtein, V.A. Fateev and \AZm}
{Quantum S-matrix of the 1+1 dimensional Toda chain}
{\PL {\bf B87} (1979) 389}

\refdef\rBCDSa\BCDSa{\dur, \PL {\bf B227} (1989) 411}

\rdef\rBCDSc\BCDSc{\dur}
{Affine Toda field theory and exact S-matrices}{\NP {\bf B338} (1990) 689}

\rdef\rBCDSe\BCDSe{\dur}
{Multiple poles and other features of affine Toda field theory}
{\NP {\bf B356} (1991) 469}

\rdef\rBSa\BSa{H.\ts W.\ts Braden and R.\ts Sasaki}
{The S-matrix coupling dependence for a, d and e affine Toda
field theory}
{\PL {\bf B255} (1991) 343}

\rdef\rCDSb\CDSb{E.\ts Corrigan, \ped\ and R.\ts Sasaki}
 {On a generalised bootstrap principle}
 {\NP {\bf B408} (1993) 579--599}

\rdef\rCDRa\CDRa{E.\ts Corrigan, \ped\ , R.H.\ts Rietdijk}
{Aspects of affine  Toda field theory on a half line}
{to appear in Proceedings of
\lq Quantum field theory, integrable models and beyond',
Yukawa Institute for Theoretical Physics, Kyoto University, February 1994;
hepth/9407148}

\rdef\rCDRSa\CDRSa{E.\ts Corrigan, \ped\ , R.H.\ts Rietdijk and R.\ts Sasaki}
{Affine Toda field theory on a half line}
{\PL {\bf B333} (1994) 83}

\rdef\rCMa\CMa{P.\ts Christe and G.\ts Mussardo}
{Integrable systems away from criticality: the Toda field theory and S
matrix of the tricritical Ising model}
{\NP {\bf B330} (1990) 465}

\rdef\rCMb\CMb{P.\ts Christe and G.\ts Mussardo}
{Elastic S-matrices in (1+1) dimensions and Toda field theories}
{\IJMP {\bf A5} (1990) 4581}

\rdef\rDf\Df{\ped }
{Root systems and purely elastic S-matrices, I \& II}
{\NP {\bf B358} (1991) 654; \NP {\bf B374} (1992) 741}

\rdef\rDi\Di{\ped}
{A remark on the coupling-dependence in affine Toda field theories}
{\PL {\bf B312} (1993) 291}

\rdef\rDGZb\DGZb{G.\ts W.\ts Delius, M.\ts T.\ts Grisaru and D.\ts Zanon}
{Quantum conserved currents in affine Toda theories}
{\NP {\bf B385} (1992) 307}

\rdef\rDGZa\DGZa{G.\ts W.\ts Delius, M.\ts T.\ts Grisaru and D.\ts Zanon}
 {Exact S-matrices for non simply-laced affine Toda theories}
 {\NP {\bf B282} (1992) 365}

\rdef\rDTWa\DTWa{R. M. DeLeonardis, S.E. Trullinger and R.F.Wallis}
{Theory of boundary effects on sine-Gordon solitons}
{\JAP {\bf 51} (1980) 1211}

\rdef\rDTWb\DTWb{R. M. DeLeonardis, S.E. Trullinger and R.F.Wallis}
{Classical excitation energies for a finite length sine-Gordon system}
{\JAP {\bf 53} (1982) 699}

\rdef\rFKc\FKc{A.\ts Fring and R.\ts K\"oberle}
{Factorized scattering in the presence of reflecting boundaries}
{\NP {\bf B421} (1994) 159}

\rdef\rFKd\FKd{A.\ts Fring and R.\ts K\"oberle}
{Affine Toda field theory in the presence of reflecting boundaries}
{\NP {\bf B419} (1994) 647}

\rdef\rFKe\FKe{A.\ts Fring and R.\ts K\"oberle}
{Boundary bound states in affine Toda field theory}
{Swansea preprint SWAT-93-94-28; hep-th/9404188}

\rdef\rMv\Mv{A. MacIntyre}
{Integrable boundary conditions for classical sine-Gordon theory}
{Durham preprint DTP-94/39; hepth/9410026}

\rdef\rNa\Na{M.\ts R.\ts Niedermaier}
{The quantum spectrum of conserved charges in affine Toda theories}
{\NP {\bf 424} (1994) 184}

\rdef\rOTb\OTb{D.\ts I.\ts Olive and N.\ts Turok}
{The Toda lattice field theory hierarchies and zero-curvature
conditions in Kac-Moody algebras}
{\NP {\bf B265} (1986) 469}

\rdef\rSk\Sk{R.\ts Sasaki}
{Reflection bootstrap equations for Toda field theory}
{Kyoto preprint YITP/U-93-33; hep-th/9311027}

\rdef\rDGZb\DGZb{G.\ts W.\ts Delius, M.\ts T.\ts Grisaru and D.\ts Zanon}
{Quantum conserved currents in affine Toda theories}
{\NP {\bf B385} (1992) 307}

\rdef\rFTa\FTa{L.D. Faddeev and L.A. Takhtajan}
{Hamiltonian methods in the theory of solitons}
{Springer Verlag 1987}

\rdef\rGd\Gd{S.\ts Ghoshal}
{Boundary state boundary $S$ matrix of the sine-Gordon model}
{{\it Int. J. Mod. Phys.} {\bf A9} (1994) 4801}

\rdef\rGZa\GZa{S.\ts Ghoshal and \AZm}
{Boundary $S$ matrix and boundary state in two-dimensional
integrable quantum
field theory}
{{\it Int. J. Mod. Phys.} {\bf A9} (1994) 3841}

\rdef\rKMa\KMa{T.\ts R.\ts Klassen and E.\ts Melzer}
{Purely elastic scattering theories and their
ultraviolet limits}
{{\NP} {\bf B338} (1990) 485}

\rdef\rLa\La{G.L. Lamb, Jr}
{Elements of Soliton Theory}
{John Wiley and Sons Inc. 1980}

\rdef\rMOPa\MOPa{A. V. Mikhailov, M. A. Olshanetsky and A. M. Perelomov}
{Two-dimensional generalised Toda lattice}
{\CMP {\bf 79} (1981) 473}

\rdef\rOTa\OTa{D. I. Olive and N. Turok}
{The symmetries of Dynkin diagrams and the reduction of Toda field equations}
 {{\it Nucl. Phys.} {\bf B215} (1983) 470}

\rdef\rOTd\OTd{D. Olive and N. Turok}
{Local conserved densities and zero curvature conditions for Toda lattice
field theories}
{\NP {\bf B257} (1985) 277}

\rdef\rSSWa\SSWa{H. Saleur, S. Skorik and N.P. Warner}
{The boundary sine-Gordon theory: classical and semi-classical analysis}
{USC-94-013; hep-th/9408004}

\rdef\rKSa\KSa{A. Kapustin and S. Skorik}
{On the non-relativistic limit of the quantum sine-Gordon model with
integrable boundary condition}
{CALT-68-1949; hep-th/9409097}

\rdef\rSk\Sk{R.\ts Sasaki}
{Reflection bootstrap equations for Toda field theory}
{in {\it Interface between Physics and Mathematics}, eds W. Nahm and J-M Shen,
(World Scientific 1994) 201}

\rdef\rSZa\SZa{R.\ts Sasaki and F.\ts P.\ts Zen}
{The affine Toda S-matrices vs perturbation theory}
{{\it Int. J. Mod. Phys.} {\bf 8} (1993) 115}

\rdef\rSl\Sl{E.\ts K.\ts Sklyanin}
{Boundary conditions for integrable equations}
{\FAP {\bf 21} (1987) 164}

\rdef\rWa\Wa{G. Wilson}
{The modified Lax and and two-dimensional Toda lattice equations
associated with simple Lie algebras}
{{\it Ergod. Th. and Dynam. Sys.} {\bf 1} (1981) 361}

\rdef\rSm\Sm{E.\ts K.\ts Sklyanin}
{Boundary conditions for integrable quantum systems}
{\JP {\bf A21} (1988) 2375}

\rdef\rTa\Ta{V.\ts O.\ts Tarasov}
{The integrable initial-value problem on a semiline:  nonlinear
\hfill\break Schr\"odinger
and sine-Gordon equations}
{{\it Inverse Problems} {\bf 7} (1991) 435}

\rightline{DTP-94/57}
\rightline{hep-th/9501098}
\bigskip
\centerline{\bf Classically integrable boundary conditions for affine Toda
field theories}
\vskip 3pc
\centerline{P. Bowcock, E. Corrigan, P.E. Dorey, R.H. Rietdijk}
\bigskip
\centerline{Department of Mathematical Sciences}
\centerline{University of Durham}
\centerline{Durham DH1 3LE, England}
\medskip
\vskip 4pc
\centerline{\bf Abstract}
\bigskip
Boundary conditions compatible with classical integrability are studied both
directly, using an approach based on the explicit construction
of conserved quantities, and indirectly by first developing a generalisation
of the Lax pair idea. The latter approach is closer to the spirit of earlier
work by Sklyanin and yields a complete set of conjectures
for permissible boundary conditions for any affine Toda field theory.
\vskip 8pc
\noindent January 1995

\vfill
\eject

\newsec{Introduction}

Besides its classical integrability
\NRF\rMOPa{\MOPa\semi\Wa\semi\OTa}
\NRF\rOTd\OTd
\NRF\rOTb\OTb\refs{\rMOPa -\rOTb}, affine Toda field theory appears to
be quantum integrable and, over recent years, much has been discovered
concerning the spectrum and scattering in the real coupling
regime~\NRF\rAFZa\AFZa
\NRF\rBCDSc{\BCDSc\semi\BCDSe}%
\NRF\rCMa{\CMa\semi\CMb}%
\NRF\rBSa{\BSa\semi\SZa}%
\NRF\rDf{\Df}%
\NRF\rDGZa{\DGZa\semi\CDSb\semi\Di}\refs{\rAFZa  -\rDGZa}.
The classical integrability of the theory stems from the existence of
a Lax pair representation of the field equations, leading to
an infinite set of independent conserved charges in involution.

More recently, there has been some interest in examining the Toda
models on a half-line or on a finite interval. In particular,
certain special solutions to
appropriately modified bootstrap relations have been obtained by
Fring and K\"oberle
\NRF\rFKc{\FKc\semi\FKd\semi\FKe}\refs{\rFKc} and by Sasaki
\NRF\rSk\Sk{\refs{\rSk}
but without establishing the
precise nature of the boundary conditions responsible for them.
For the special case of the sine (or sinh)-Gordon model, there is already
a substantial literature, examining the classically permissible boundary
conditions which preserve integrability and calculating the effects of the
allowable boundaries in quantum field theory
\NRF\rDTWa{\DTWa\semi\DTWb}
\NRF\rSl{\Sl\semi\Sm}
\NRF\rTa\Ta}
\NRF\rGZa\GZa
\NRF\rSSWa{\SSWa\semi\KSa}
\NRF\rGd\Gd
\NRF\rMv\Mv
\refs{\rDTWa -\rMv}.
It appears \refs{\rGZa}
the most general
boundary condition (at $x^1=0$, say) is of the form:
\eqn\sgboundary{{\partial\phi\over\partial x^1}={a\over\beta}\sin \beta
\left({\phi -\phi_0 \over 2}\right)\qquad \hbox{at}\qquad x^1=0,}
where $a$ and $\phi_0$ are arbitrary constants, and $\beta$
is the sine-Gordon coupling. Some years ago, Sklyanin and others
\refs{\rSl ,\rTa}
have argued for restricted versions
of \sgboundary\  in the classical theory, and MacIntyre
\refs{\rMv}
has recently demonstrated that \sgboundary\  preserves classical
integrability. Ghoshal and Zamolodchikov
\refs{\rGZa ,\rGd}
have also presented arguments supporting the idea that the quantum sine-Gordon
theory with a
boundary does indeed have a pair of additional coupling constants associated
with the boundary, although it is not yet completely clear in what way
these are
related to the parameters
$a$ and $\phi_0$ appearing in \sgboundary\ (but see also
\refs{\rSSWa}
for further developments in this direction).

In an attempt to explain the  variety of solutions exhibited
in \refs{\rFKc ,\rSk}, the classical
charges of affine Toda theory were studied in
\NRF\rCDRSa\CDRSa\NRF\rCDRa\CDRa\refs{\rCDRSa ,\rCDRa}. Surprisingly,
it was discovered that the possible boundary conditions for the
$a_n^{(1)}$ and
$d_n^{(1)}$ theories are highly constrained by the requirement that
there should be
conserved modifications of the spin two or three charges even in the presence
of
the boundary. Effectively, in those cases, there is only a discrete ambiguity
and the possible boundary conditions are summarised by adding a term to the
action\foot{The notation and conventions for affine Toda field theory
are those of \refs{\rBCDSc}} of the form
\eqn\baction{{\cal L}_{\rm boundary}=-\delta (x^1) {\cal B}(\phi ),}
where
\eqn\allboundary{{\cal B}={m\over \beta^2}\sum_0^rA_ie^{{\beta\over 2}
\alpha_i\cdot\phi},}
and the coefficients $A_i,\ i=0,\dots ,r$ are a set of real numbers
with
\eqn\gboundary{\hbox{\bf either}\ |A_i|=2\sqrt{n_i}, \ \hbox{for}
\ i=0,\dots ,r\
\hbox{\bf or}\ A_i=0\ \hbox{for}\ i=0,\dots ,r\ .}
In this paper, these arguments will be elaborated and extended to include spin
four charges (thereby encompassing the the case $e_6^{(1)}$).

The constraints on the boundary term are seen to be necessary,
following the arguments of \refs{\rCDRSa ,\rCDRa} and below,
but not sufficient; there is always the
chance that a study of charges with spins greater than three or four may lead
to stronger constraints on the coefficients $A_i$ appearing in eq\allboundary .
With this worry in mind, it is clearly desirable to find an
alternative approach to the classical integrability, preferably one
which is
close to the Lax pair idea, even in the presence of boundary conditions
at one  specific value of $x^1$. (Or possibly two, if the theory is defined on
an interval.)

One of the purposes of this article is to report on a definition of the Lax
pair
for affine Toda theory which successfully incorporates
as a consequence
of the zero curvature requirement
not only the
equations of motion but also the boundary conditions.
The Lax pair argument provides the
missing sufficiency requirement for the boundary condition restrictions
and permits conjectures to be made for possible boundary conditions
in those cases for which the direct approach in terms of charges is not
tractable.

One corollary of being able to do
this is the discovery that although the form \allboundary\
appears to be universal, the further
constraints \gboundary\ are not universal. In fact,
the stringent constraints on the coefficients $A_i$ are peculiar to
the simply-laced Toda field theories. The non-simply laced theories
are curiously different; for them,  some of the coefficients can be
chosen freely\foot{These observations may be connected with the remarkable
duality properties of quantum affine Toda field theory on the full line:
the simply-laced
data (and $a_{2n}^{(2)}$) lead to self-dual field theories while the
non-simply-laced data (except $a_{2n}^{(2)}$) are really dual pairs
\NRF\rDGZa{\DGZa\semi\CDSb}\refs{\rDGZa}.}.
The $a_{2}^{(2)}$ example illustrates this. An examination of the spin five
charge in the presence of a boundary condition at
$x^1=0$ leads to
a boundary condition of the form \allboundary , ie
$${\cal B}=A_1 e^{\phi} +A_0 e^{-\phi /2},$$
and the further constraint
$$A_0(A_1^2-2)=0.$$
The sine-Gordon theory remains the only example within the set of Toda  field
theories for which integrability dictates the form of the boundary condition
but places no further constraints on the   the boundary
coefficients\foot{See also the discussion of classical
sinh-Gordon in \refs{\rCDRa}.}.

\newsec{Conserved charges on the whole and half-line}

The usual lagrangian density for the full line Toda theory is
\eqn\ltoda{{\cal L}_T={1\over 2}
\partial_\mu\phi^a\partial^\mu\phi^a-V(\phi )}
where
\eqn\vtoda{V(\phi )={m^2\over
\beta^2}\sum_0^rn_ie^{\beta\alpha_i\cdot\phi}.}
(For  the classical discussion, the coupling constant $\beta$
and mass scale $m$
will be discarded from now on.)

On a half line, $x^1<a$, say, with a boundary condition at $x^1=a$, the
classical
lagrangian is effectively replaced by
\eqn\lhtoda{{\cal L}= \theta (-x^1 +a) {\cal L}_T - \delta (x^1 -a) {\cal B},}
where for the purposes of the present discussion ${\cal B}$ is a function of
the field only, not its derivatives. The lagrangian ${\cal L}$
leads to the field equations
on $x^1<a$ and to the boundary condition
\eqn\boundary{\partial_1\phi = -{\partial {\cal B}\over \partial\phi},}
at $x^1 =a$.

The classical conserved charges $Q_{s}$ of affine Toda field theory on the full
line can be calculated in principle using a Lax pair.
Later,  a generalisation of that procedure to the
half-line will be given but, in this section, the spin
2, 3 and 4 charges for the simply-laced cases both on the whole and the
half-line will be calculated in a pedestrian fashion,
following \refs{\rCDRSa ,\rCDRa}.

On the full line, a density $T_{\pm (s+1)}$ defines a conserved
charge $Q_{\pm s}$ if it satisfies (using the equations of motion)
\eqn\currenta{\partial_{\mp}T_{\pm (s+1)}=\d \Theta_{\pm (s-1)}}
for some $\Theta_{\pm (s-1)}$. The conserved charge is then given by
\eqn\spins{Q_{\pm s}=\int_{-\infty}^{+\infty}dx^1
(T_{\pm (s+1)}-\Theta_{\pm (s-1)}).}
For an affine Toda theory on the full line based on an algebra $g$ there is an
infinite set of conserved charges, one for each integer
$s$ of the form $s=m+nh$, where $m$ is an exponent of the
algebra $g$, $n$ is an integer, and $h$ is its Coxeter number.

In order for the theory on the half-line to
remain integrable, an infinite set of conserved charges should continue to
exist.
Clearly, translational invariance is destroyed and
the only combination of spin $\pm 1$ charges which is
left in the theory restricted to a half
line is the energy. For any boundary condition this is given by
\eqn\energy{E=\hat{Q}_{1} + \hat{Q}_{-1} + \B,}
where hatted quantities are the standard
densities integrated over the half line. For the higher spins
considered here
it is found there are conserved quantities of the form
\eqn\pcharge{P_{s} = \hat Q_{s} + \hat Q_{-s} - \Sigma_{s},}
where the boundary term $\Sigma_{s}$ is defined by the condition
\eqn\tboundary{(T_{s+1} + \Theta_{s-1} - T_{-s-1} - \Theta_{-s+1})
= \partial_0\Sigma_{s}.}
Such conserved quantities will be referred to as
\lq spin $s$'.

For $s=1$,  condition \pcharge\ is automatically fulfilled by choosing
$\Sigma_{1} = - \B$. However, for the higher spin charges \tboundary\
severely restricts the
form of the boundary conditions. It will be shown that $\B$ must take the form
\allboundary\
where the coefficients $A_i, i=0,\ldots,n$ are a set of real numbers.
Furthermore, conservation of the charges considered here restricts the values
of
the $A_i$ in the cases $e_{6}^{(1)}, d_{n}^{(1)}, a_{n}^{(1)}, n \geq 2$ to
those satisfying \gboundary .

\subsec{Spin two charges}

A general ansatz for $T_{\pm 3}$ (using light-cone coordinates
$x^{\pm}=(x^0\pm x^1)/\surd 2$) reads
\eqn\tthree{T_{\pm 3} = {1\over 3} A_{abc} \d \phi_a \d \phi_b \d \phi_c +
B_{ab} \d^2 \phi_a \d \phi_b,}
where the coefficients $A_{abc}$ are completely symmetric and the coefficients
$B_{ab}$ are antisymmetric. An explicit calculation reveals that \currenta\ is
satisfied for
\eqn\thone{\Theta_{\pm 1}=-{1\over 2} B_{ab}\d\phi_aV_b,}
provided that
\eqn\aba{A_{abc}V_a+B_{ab}V_{ac}+B_{ac}V_{ab}=0,}
where
\eqn\notation{V_b={\partial V\over \partial\phi_b},\qquad V_{bc}=
{\partial^2 V\over \partial\phi_b\partial\phi_c},\qquad\hbox{etc.}}

For practical calculations it is convenient to introduce the notation
\eqn\abdef{A_{ijk}=A_{abc}(\alpha_i)_a(\alpha_j)_b(\alpha_k)_c,\qquad
B_{ij}=B_{ab}(\alpha_i)_a(\alpha_j)_b,}
and
$$C_{ij}=\alpha_i\cdot\alpha_j.$$
Then eq\aba\ implies
\eqn\abb{ A_{ijk}+B_{ij}C_{ik}+B_{ik}C_{ij}=0.}
This equation is very restrictive and fixes both $A_{ijk}$ and $B_{ij}$ up to
an overall constant. In fact,
$B_{ij}$ is non-zero only for the $a_n^{(1)}$ cases and, for
those cases (and $n>1$), $B_{ij}=0$ except for $j=i\pm 1\ {\rm mod} \ n+1$, and
$B_{i-1\, i}=B_{i\, i+1},\ i=1,\dots , n+1$. The sets
of coefficients $A_{abc}$ and $B_{ab}$ are found
by inverting the transformations  \abdef\ and lead to the following
conserved current densities for $a_{n}^{(1)} \; (n>1)$:
\eqn\tthreean{T_{\pm 3}^{(n)}={2(2i) \over 3 \sqrt{n+1}}
\delta_{a+b+c,0\ {\rm mod}\ (n+1)}\d \phi_{a} \d \phi_{b} \d \phi_{c} +
{\Gamma_2^a}
\delta_{a+b,n+1} \d^{2} \phi_a \d \phi_b,}
where
\eqn\Gam{\Gamma_s^a =
{\sin \left( {sa \pi \over h} \right) \over \sin^s \left( {a \pi \over h}
\right)} ,}
with $h$ the Coxeter number of the algebra.

On the half-line, a spin two charge $P_{2}$ exists if condition \tboundary\ is
satisfied for some $\Sigma_{2}$. Substituting the definitions \tthree\ and
\thone\ in \tboundary\ for $s=2$ it is found that $\Sigma_2$ does not exist
unless the following two conditions hold at $x^1=0$:
\eqn\conditiona{A_{abc}{\cal B}_a+2B_{ab}{\cal B}_{ac}
+2B_{ac}{\cal B}_{ab}=0,}
\eqn\conditionb{{1\over 3}A_{abc}{\cal B}_a{\cal B}_b{\cal B}_c+2B_{ab}
V_a{\cal B}_b=0.}
Similar notation to that of \notation\ is being used for derivatives of $\B$.
Once conditions \conditiona , \conditionb\ are solved the extra piece in
eq\pcharge\ will be given by
\eqn\sitwo{\Sigma_{2} = - \sqrt{2} B_{ab} \partial_0 \phi_a \B_b.}
Both conditions involve the boundary term $\B$. Comparing \conditiona\ with
\aba\ reveals that $\B$ must be equal to
$$\sum_0^rA_ie^{\alpha_i\cdot\phi /2},$$
apart from an additive arbitrary constant. The second condition,
eq\conditionb , is nonlinear in the boundary term and therefore provides
equations for the constant coefficients $A_i$ in terms of the coefficients in
the potential. In one way of analysing these equations, the explicit
expressions for $A_{abc}$ and $B_{ab}$, which for $a_{n}^{(1)}$ are defined by
\tthreean , are substituted to find that the $A_i$ have to satisfy \boundary .

\subsec{Spin three charges}

In this case, the following ansatz for the conserved current is appropriate:
\eqn\tfour{T_{\pm 4} = {1 \over 4} A_{abcd} \d \phi_a \d \phi_b \d \phi_c \d
\phi_d + {1 \over 2} B_{abc} \d^2 \phi_a \d \phi_b \d \phi_c +
{1 \over 2} D_{ab} \d^3 \phi_a \d \phi_b,}
where $A_{abcd}$ and $D_{ab}$ are completely symmetric. The
coefficients $B_{abc}$ are
symmetric in their last two indices but are ambiguous up to a
totally symmetric part; adding a totally
symmetric part to $B_{abc}$ will only add a total $\d$ derivative to
$T_{\pm 4}$, which will not change condition \currenta.

The above expression
corresponds to a conserved charge on the whole line if
\eqn\abca{\eqalign{&B_{[ab]c}V_c-D_{c[a}V_{b]c}=0\cr
&A_{abcd}V_a+{1\over 2}B_{a(bc}V_{d)a}-{1\over 2}V_{a(d}B_{bc)a}+
{1\over 2}D_{a(b}V_{cd)a}=0.\cr}}
Then
\eqn\thtwo{\Theta_{\pm 2}=-{1\over 4}B_{abc}V_b\d\phi_a\d\phi_c-{1\over 4}
D_{ab}V_a\d^2\phi_b.}
Here it has been convenient to introduce a bracket notation
\eqn\sym{\eqalign{
&M_{(a_1 \cdots a_n)}={1 \over n!} \sum_{\{\rm permutations\}}
M_{p(a_1) \cdots p(a_n)} , \cr
&M_{[a_1 \cdots a_n]}={1 \over n!} \sum_{\{\rm permutations\}}
{\rm sign}\{p\} M_{p(a_1) \cdots p(a_n)} . \cr}}
Eqs\abca\ have solutions for both $a_{n}^{(1)}$ and $d_{n}^{(1)}$. They have
been computed using Mathematica after transforming the equation into a form
similar to \abb . The expressions for $T_{\pm 4}$ which are found this way can
be written in a nice form by choosing the completely symmetric part of
$B_{abc}$ conveniently. For $a_{n}^{(1)}$ the expressions are
\eqn\tfouran{\eqalign{
T_{\pm 4}^{(n)} = & {2(2i)^2 \over 4(n+1)}
\left\{
\delta_{a+b+c+d,0\ {\rm mod}\ (n+1)}
- {3 \over 2} \delta_{a+b,n+1} \delta_{c+d,n+1}
\right\}
\cr
& \;\;\;\;\;\;\;\;\;\;\;\;\;\;\;\;\;\;\;\;\;\;\;\;\;\;\;\;\;\;\;\;\;\;\;\;\;\;
\;\;\;\;\;\;\;\;\;\;\;\;\;\;\;\;\;\;\;\;
\times \d \phi_a \d \phi_b \d \phi_c \d \phi_d \cr
& + {3(2i) \Gamma_2^a \over 2  \sqrt{n+1}}
\delta_{a+b+c,0\ {\rm mod}\ (n+1)}
\d^2 \phi_a \d \phi_b \d \phi_c \cr
& + {\Gamma_3^a}
\delta_{a+b,0\ {\rm mod}\ (n+1)}
\d^3 \phi_a \d \phi_b, \cr
}}
while for $d_{n}^{(1)}$
\eqn\tfourdne{\eqalign{
T_{\pm 4}^{(n)} =
& -{1 \over (n-1)}
\left\{
\delta_{a+b+c+d,2(n-1)} +3  \delta_{a+b-c-d,0} - 4 \delta_{-a+b+c+d,0\ {\rm
mod}\ 2(n-1)}
- 3 \delta_{a,b} \delta_{c,d}
\right\}
\cr
& \;\;\;\;\;\;\;\;\;\;\;\;\;\;\;\;\;\;\;\;\;\;\;\;\;\;\;\;\;\;\;\;\;\;\;\;\;\;
\;\;\;\;\;\;\;\;\;\;\;\;\;\;\;\;\;\;\;\; \times
\d \phi_a \d \phi_b \d \phi_c \d \phi_d  \cr
& + {3 \Gamma_2^a \over \sqrt{2(n-1)}}
\Biggl[
\left\{
\delta_{a+b+c,2(n-1)} - 2 \delta_{a-b+c,0} + \delta_{-a+b+c,0}
\right\}
\d^2 \phi_a \d \phi_b \d \phi_c  \cr
& - \left[ 1 + (-)^{n+a} \right]
\left\{
\left( \d \phi_{s} \right)^2 + \left( \d \phi_{s'} \right)^2
\right\}
\d^2 \phi_a
- 2 \left[ 1 - (-)^{n+a} \right]
\d \phi_{s} \d \phi_{s'} \d^2 \phi_a
\Biggr] \cr
& + {\Gamma_3^a }
\delta_{a,b} \d^3 \phi_a \d \phi_b -4
\Delta_{\pm 4}^{(n)}, \cr
}}
with
\eqn\deltae{\eqalign{
& \Delta_{\pm 4}^{(n)} =
 \left\{
\d \phi_{s} \d^3 \phi_{s} + \d \phi_{s'} \d^3 \phi_{s'}
\right\}
- {3 \over 2(n-1)}
\left\{ \left( \d \phi_{s} \right)^2 +
\left( \d \phi_{s'} \right)^2
\right\}
\left( \d \phi_a \right)^2 \cr
& \;\;\;\;\; + {(n-4) \over 4(n-1)}
\left\{
\left( \d \phi_{s} \right)^4 +
\left( \d \phi_{s'} \right)^4
\right\}
+ {3(n-2) \over 2(n-1)}
\left( \d \phi_{s} \right)^2 \left( \d \phi_{s'} \right)^2 \cr
}}
for even $n$, and
\eqn\deltao{\eqalign{
& \Delta_{\pm 4}^{(n)} =
\left\{
\d \phi_{s} \d^3 \phi_{\bar s} + \d \phi_{\bar s} \d^3 \phi_{s}
\right\}
- {3 \over (n-1)}
\d \phi_{s} \d \phi_{\bar s} \left( \d \phi_a \right)^2 \cr
& \;\;\;\;\; + {1 \over 4} \left\{ \left( \d \phi_{s} \right)^4 +
\left( \d \phi_{\bar s} \right)^4 \right\} +
{3(n-3) \over 2(n-1)}
\left( \d \phi_{s} \right)^2 \left( \d \phi_{\bar s} \right)^2 \cr
}}
for odd $n$. For $d_{4}^{(1)}$ there is an extra spin 3 charge which is defined
by
\eqn\tdfour{\eqalign{T_{\pm 4}^{(4)}= & {1 \over 2}
\left\{ \left( \d \phi_{1} \right)^2 - \left( \d \phi_{2} \right)^2 \right\}
\left\{ \left( \d \phi_{s} \right)^2 - \left( \d \phi_{s'} \right)^2 \right\} +
 \left\{
\d \phi_{s} \d^3 \phi_{s} - \d \phi_{s'} \d^3 \phi_{s'} \right\} \cr
& - {\sqrt{2}} \left\{ \d^2 \phi_{s}
\left[ \d \phi_{1} \d \phi_{s'} + \d \phi_{2} \d \phi_{s} \right]
- \d^2 \phi_{s'}
\left[ \d \phi_{1} \d \phi_{s} + \d \phi_{2} \d \phi_{s'} \right] \right\}.
\cr }}

The conditions for a spin 3 charge $P_{3}$ to exist on the half-line are given
by \tboundary\ for $s=3$. One finds that at $x_1=0$ the following equations
have to be satisfied:
\eqn\abcb{\eqalign{&B_{[ab]c}{\cal B}_c-2D_{c[a}{\cal B}_{b]c}=0,\cr
&A_{abcd}{\cal B}_a+B_{a(bc}\B_{d)a}-\B_{a(d}B_{bc)a}+
2D_{a(b}\B_{cd)a}=0,\cr
& - {1 \over 2} \left( A_{abcd} \B_a \B_b \B_c + B_{abc} \B_{ad} \B_b \B_c
+ B_{abd} V_a \B_b - B_{(db)a} V_a \B_b  \right. \cr
& \left. \;\;\;\;\;\;\;\;\;\;\;\;\;\;\;\;\;\;\;\;
+ D_{a(d} V_{b)a} \B_{b} - 2 D_{ab} V_a \B_{bd} \right)
= {\partial \Sigma_{3}^{(0)} \over \partial \phi_d},}}
and an expression for  $\Sigma_{3}^{(0)}$ determined.
Once the latter is found,
the additional piece in the conserved quantity is given by
\eqn\sithree{\Sigma_{3} = - {1 \over 2} B_{abc} \partial_0 \phi_a \partial_0
\phi_b \B_c - D_{ab} \B_a \partial_{0}^{2} \phi_b
- {1 \over 2} D_{ab} V_a \B_b + \Sigma_{3}^{(0)}.}
The first two of conditions \abcb\ are automatically satisfied by the general
boundary term ${\cal B}$ given in \allboundary\ as a consequence of \abca. The
last condition is
non-linear in the boundary term and gives conditions on the parameters $A_i$.
For $a_{n}^{(1)}$ these are the same conditions as those found from the spin
two charge. The existence of a conserved spin three charge also restricts the
parameters for $d_{n}^{(1)}$ to satisfy \gboundary. Then $\Sigma_{3}^{(0)}$ is
given by
\eqn\sizeroa{
\Sigma_{3}^{(0)} = \sum_{i,j=0}^{n}
A_{i}^{2} A_{j}
\left[
{1 \over 6} \delta_{ij} - {1 \over 4} \A_{ij}
\right]
e^{ \left( 2 \alpha_{i} + \alpha_{j} \right) \cdot \phi /2}
}
for $a_{n}^{(1)}$ and
\eqn\sizerod{\eqalign{
\Sigma_{3}^{(0)} =
& \sum_{i,j=0}^{n}
\left\{
A_{i}^{2} A_{j}
\left[
- {1 \over 24} \delta_{ij} + {1 \over 16} \A_{ij} +
{3 \over 16}
\left(
\delta_{i,0} \delta_{j,1} + \delta_{i,1} \delta_{j,0} +
\delta_{i,n-1} \delta_{j,n} + \delta_{i,n} \delta_{j,n-1}
\right)
 \right. \right. \cr
& \left. \left. \;\;\;\;\;\;\;\;\;\;\;\;\;\;\;\;\;\;\;\;\;\;\;\;\;
- {1 \over 16}
\left(
\delta_{i,0} \delta_{j,0} + \delta_{i,1} \delta_{j,1} +
\delta_{i,n-1} \delta_{j,n-1} + \delta_{i,n} \delta_{j,n}
\right)
\right]
e^{ \left( 2 \alpha_{i} + \alpha_{j} \right) \cdot \phi /2}
\right\}
\cr
& - {3 \over 8}
\left\{
A_0 A_1 A_2
e^{
\left( \alpha_0 + \alpha_1 + \alpha_2 \right) \cdot \phi /2} +
A_{n-2} A_{n-1} A_{n}
e^{
\left( \alpha_{n-2} + \alpha_{n-1} + \alpha_{n} \right) \cdot \phi /2}
\right\} \cr
}}
for $d_{n}^{(1)}$. Here $\A^{ij}$ is the adjacency matrix
$2 \delta_{ij} - C^{ij}$. For the extra spin 3 charge of $d_{4}^{(1)}$ which
is defined by \tdfour , $\Sigma_{3}^{(0)}$ vanishes.

\subsec{Spin four charges}

The most general ansatz for $T_{\pm 5}$ reads
\eqn\Tfive{\eqalign{
T_{\pm 5} = & {1 \over5}
A_{abcde}\d\phi_a\d\phi_b\d\phi_c\d\phi_d\d\phi_e
+{1\over 3}B_{abcd}\d^2\phi_a\d\phi_b\d\phi_c\d\phi_d  \cr
& + {1 \over 2}D_{abc} \d^2\phi_a\d^2\phi_b\d\phi_c
+E_{ab} \d^4 \phi_a \d \phi_b, \cr
}}
with $A_{abcde}$ is completely symmetric, $B_{abcd}$ symmetric in all but its
first index and only defined modulo a totally symmetric part.
The set of coefficients $D_{abc}$ is
symmetric in its first two indices and $E_{ab}$ is anti-symmetric. The above
ansatz for $T_{\pm 5}$ corresponds to a conserved charge on the whole line
under the following constraints
\eqn\abcda{\eqalign{
& D_{abc}V_{c} - 4E_{c(a} V_{b)c} = 0, \cr
& B_{bcda} V_{a} - B_{(cd)ab} V_{a} + D_{ab(c} V_{d)a} -
{1 \over 2} V_{a(c} D_{d)ab} - {1 \over 2} D_{a(cd)}V_{ab} \cr
& \ \ \ \ \ \ \ \ \ \
- 2E_{ab} V_{acd} + 2E_{a(c}V_{d)ab} = 0, \cr
& A_{abcde}V_{a} + {1 \over 3} B_{a(bcd} V_{e)a}
- {1 \over 3} V_{a(e} B_{bcd)a} - {1 \over 3} D_{a(bc} V_{de)a} +
{2 \over 3} E_{a(b} V_{cde)a} = 0.\cr
}}
Then $\Theta_{\pm 3}$ is given by
\eqn\Ththree{\eqalign{
\Theta_{\pm 3} =
& - {1 \over 6} \left\{ B_{bcda} V_{a} + D_{abc}V_{ad} + E_{ab}V_{acd} \right\}
\d\phi_b\d\phi_c\d\phi_d  \cr
& + {1 \over 2} \left\{ E_{ab}V_{bc} + E_{cb}V_{ba} \right\}
\d^2\phi_a\d\phi_c - {1 \over 2} E_{ab} V_b \d^3 \phi_a.
}}
Eqs \abcda\ only have non-trivial solutions for $a_{n}^{(1)}$, $d_{5}^{(1)}$
and $e_{6}^{(1)}$. For $a_{n}^{(1)}$ these define the following conserved
currents
\eqn\Tfivean{\eqalign{T_{\pm 5}^{(1)}  =&
{2(2i)^3 \over 5(n+1)^{{3 \over 2}}}
\left\{ \delta_{a+b+c+d+e, 0\ {\rm mod}\ (n+1)}
- (10 / 3) \delta_{a+b,n+1} \delta_{c+d+e ,0\ {\rm mod}\ (n+1)}
\right\} \cr
& \ \ \ \ \ \ \ \ \ \ \ \ \ \ \ \ \ \ \ \ \ \ \ \ \ \ \ \ \ \ \ \ \ \ \ \ \ \
\ \ \ \ \ \ \ \ \ \ \ \ \ \ \ \ \ \ \ \
\times \d \phi_a \d \phi_b \d \phi_c \d \phi_d \d \phi_e  \cr
& + {(2i)^2 \over  (n+1)}
\Biggl[ (4 / 3) \Gamma_2^a
\left[ \delta_{a+b+c+d,0\ {\rm mod}\ (n+1)}
- (3 / 2) \delta_{a+b,n+1} \delta_{c+d,n+1}
\right]  \cr
& \ \ \ \ \ \ \ \ \ \ \ \ \ \ \  + \Gamma_2^{a+b}
\left[ \delta_{a+b+c+d,0\ {\rm mod}\ (n+1)} -
\delta_{a+b,n+1} \delta_{c+d,n+1}
\right]
\Biggr] \cr
& \ \ \ \ \ \ \ \ \ \ \ \ \ \ \ \ \ \ \ \ \ \ \ \ \ \ \ \ \ \ \ \ \ \ \ \ \ \
\ \ \ \ \ \ \ \ \ \ \ \ \ \ \ \ \ \ \ \
\times \d^2 \phi_a \d \phi_b \d \phi_c \d \phi_d  \cr
& + {(2i) \over  (n+1)^{{1 \over 2}}}
\left\{ - {4 \over 3} \Gamma_3^a - {4 \over 3} \Gamma_3^b
+ {4 \over 3} + \Gamma_2^a \Gamma_2^b \right\} \delta_{a+b+c, 0\ {\rm mod}\
(n+1)}
\d^2 \phi_a \d^2 \phi_b \d \phi_c  \cr
& + {\Gamma_4^a } \delta_{a+b,n+1}
\d^4 \phi_a \d \phi_b.
}}
For $d_{5}^{(1)}$ and $e_{6}^{(1)}$ the expressions for $T_{\pm 5}$ are quite
long. Appendix A contains further details on these cases.

For a spin four charge to exist on the half line the condition \tboundary\ must
hold.
This requirement leads to the conditions
\eqn\abcdb{\eqalign{
&D_{abc}{\cal B}_{c} - 8 E_{c(a}{\cal B}_{b)c} = 0, \cr
&B_{bcda}{\cal B}_{a} - B_{(cd)ab}{\cal B}_{a} +
2D_{ab(c}{\cal B}_{d)a} -
\B_{a(c}D_{d)ab} - D_{a(cd)}\B_{ab} \cr
&\ \ \ \ \
- 8E_{ab}{\cal B}_{acd}
+ 8E_{a(c}{\cal B}_{d)ab} = 0, \cr
&A_{abcde}{\cal B}_{a} -
{2 \over 3}{\cal B}_{a(e}B_{bcd)a} +
{2 \over 3}B_{a(bcd}{\cal B}_{e)a} -
{4 \over 3} D_{a(bc}{\cal B}_{de)a} +
{16 \over 3} E_{a(b}{\cal B}_{cde)a} = 0,\cr}}
and
\eqn\abcdbb{\eqalign{&A_{abcde}\B_c\B_d\B_e -
\B_{c(b}B_{a)cde}\B_d\B_e + B_{cde(a}\B_{b)c}\B_d\B_e +
D_{cde}\B_{ac}\B_{bd}\B_e \cr
&\ \ \ \ \ + {1 \over 2}B_{cdab}V_c\B_d -
{1 \over 3}B_{(ab)cd}V_c\B_d -
{1 \over 6}B_{dabc}V_c\B_d -
{7 \over 6}V_{d(b}D_{a)dc}\B_c  \cr
&\ \ \ \ \ - \B_{c(a}D_{b)dc}V_d
+ D_{cd(a}\B_{b)d}V_c
- 6E_{cd}V_{c(a}\B_{b)d} +
{8 \over 3} E_{d(a}V_{b)cd}\B_c \cr
&\ \ \ \ \ + 2E_{d(a}\B_{b)c}V_{cd}
- {1 \over 6}D_{cd(a}V_{b)c}\B_d +
{1 \over 3} E_{cd}V_{abc}\B_d - {1 \over 6} D_{c(ab)}V_{cd}\B_d = 0,\cr
&{1 \over 5}A_{abcde}\B_a\B_b\B_c\B_d\B_e +
{1 \over 3}B_{abcd}V_a\B_b\B_c\B_d -
{1 \over 3}B_{bcda}V_a\B_b\B_c\B_d \cr
&\ \ \ \ \ - {1 \over 3} D_{abc}V_{ad}\B_b\B_c\B_d
+ {2 \over 3} E_{ab}V_{acd}\B_b\B_c\B_d +
{1 \over 2} D_{abc}V_a V_b \B_c \cr
&\ \ \ \ \ - 2 E_{ab}V_{ac} V_b \B_c = 0. \cr}}
Then
\eqn\Thfour{\eqalign{\Sigma_4 = &- {1 \over 3\sqrt{2}}
\left\{
B_{bcda}\B_a + 2 D_{abc}\B_{ad}
\right\} \partial_0 \phi_b \partial_0 \phi_c \partial_0 \phi_d \cr
&- {1 \over \sqrt{2}}
\left\{
{1 \over 3} B_{abcd}\B_b\B_c\B_d + D_{abc}V_b\B_c + E_{ab}V_{bc}\B_c -
E_{cb}V_{ba}\B_c - 2E_{cb}V_b\B_{ac}
\right\}\partial_0 \phi_a. }}
The first equations \abcdb\ are automatically solved when
$\B(\phi)$ has the form given in \allboundary, as a consequence of \abcda.
The other two  equations \abcdbb\ are non-linear in the boundary term
and will restrict the $A_i$ parameters in the boundary condition. The result
is consistent with what was found already for $a_{n}^{(1)}$ and $d_{5}^{(1)}$
from the spin two and three charges. To have a conserved spin four charge for
$e_{6}^{(1)}$ the boundary parameters are restricted to satisfy \gboundary,
too.

It is worth noting the following. The expressions  found for the
spin two, three and four charges for the $a_{n}^{(1)}$ theory on the
full line are in agreement with
Niedermaier's results in
\NRF\rNa\Na\refs{\rNa}. In \refs{\rNa}, expressions are given for the two
index tensor that appears in a current of arbitrary spin $s+1$ for
$a_{n}^{(1)}$:
\eqn\niedeen{\Delta T_{\pm(s+1)} = B_{ab} \d^s \phi_a \d \phi_b.}
It is found that $B_{ab}$ is restricted to have the form
\eqn\niedtwee{B_{ab} \sim \delta_{a+b, n+1} \Gamma^a_{s}.}
It seems that this result generalises to other algebras. On the basis of the
results presented in this section, one might conjecture
\eqn\conjecture{B_{ab} \sim \delta_{(a,b)}
{\gamma_s^a \over \left( \gamma_1^a \right)^s},}
where $\gamma_s$ is the $s^{th}$ eigenvector of the
Cartan matrix corresponding
to the algebra. The symbol $\delta_{(a,b)}$ means $\delta_{a,b}$ if the basis
of simple roots is real, as is
appropriate for Toda field theories with no mass degenerate conjugate pairs of
particles. Otherwise,  if the basis is (partly) complex, $B_{ab}$ becomes
(partly) anti-diagonal. This happens for $a_{n}^{(1)}$, where $\delta_{(a,b)} =
\delta_{a+b,n+1}$. For $d_{\rm odd}^{(1)}$ the basis is complex in the
$(s,\bar{s})$-subspace. Then $\delta_{(a,b)}=\delta_{a,b}$ for
$a,b=1,\ldots,n-2$, $\delta_{(s,s)}=\delta_{(\bar{s},\bar{s})} = 0$ and
$\delta_{s,\bar{s}}=1$. In fact, one would expect to
obtain \conjecture\ from the
quantum result, given that the quantum charges have eigenvalues which can be
regarded
as the components of the Cartan matrix for $g$
\NRF\rKMa\KMa\refs{\rKMa}
and taking the classical limit. (See  ref\NRF\rDGZb\DGZb\refs{\rDGZb} for
perturbative calculations of  quantum charges.)

\newsec{A Lax pair for the theory on a half line}

To establish notation, the by now standard Lax pair for the affine Toda theory
will be written
in the form
\eqn\laxfull{\eqalign{&a_0=H\cdot\partial_1\phi /2+\sum_0^r
\sqrt{m_i}(\lambda E_{\alpha_i}-1/\lambda \ E_{-\alpha_i}) e^{\alpha_i\cdot\phi
/2}\cr
&a_1=H\cdot\partial_0\phi /2+\sum_0^r
\sqrt{m_i}(\lambda E_{\alpha_i}+1/\lambda \ E_{-\alpha_i}) e^{\alpha_i\cdot\phi
/2},\cr}}
where $H, E_{\alpha_i}$ and $E_{-\alpha_i}$ are the Cartan subalgebra and the
generators
corresponding to the simple roots, respectively, of a simple Lie algebra of
rank $r$. Included in
the set of \lq simple' roots is the extra (affine) root, denoted $\alpha_0$,
which
satisfies
$$\sum_0^r\ n_i \alpha_i=0 \qquad n_0=1.$$
The coefficients $m_i$ are related to the $n_i$ by $m_i=n_i \alpha_i^2/8$.
The conjugation properties of the generators are chosen so that
\eqn\conj{a_1^\dagger  (x,\lambda )=a_1  (x,1/\lambda )
\qquad a_0^\dagger  (x,\lambda )
=a_0 (x,-1/\lambda ).}
Using the Lie algebra relations
$$[H, E_{\pm\alpha_i}]=\pm\, \alpha_i\, E_{\pm \alpha_i}\qquad
[E_{\alpha_i},E_{-\alpha_i}]=
2\alpha_i\cdot H/(\alpha_i^2),$$
the zero curvature condition for \laxfull\
$$\partial_0a_1-\partial_1a_0 +[a_0,a_1]=0$$
leads to the affine Toda field equations:
\eqn\todafull{\partial^2\phi =-\sum_0^r n_i \alpha_i e^{\alpha_i\cdot\phi}.}

What is required is a Lax pair whose zero curvature condition automatically
implies the Toda field equations on $x^1<a$ and the boundary condition at
$x^1=a$.
Merely restricting the old Lax pair to the half line will not do:
a better strategy lies
in using a Lax pair for the full line together with a
\lq reflection principle'.  A suitable modification of the Lax pair idea
will be presented below.

One of the attractive features of the Lax pair representation \laxfull\ is its
r\^ole in the generation of conserved quantities. The path-ordered exponential
\eqn\pathexp{U(x_1,x_2;\lambda )={\rm P}\exp \int_{x_1}^{x_2}\, a_1 dx^1,}
satisfies
$${d\ \over dt}U(x_1,x_2;\lambda )=U(x_1,x_2;\lambda )a_0(x_2)-a_0(x_1)
U(x_1,x_2;\lambda ).$$
Hence, provided the fields $\phi$ and their derivatives satisfy suitable
conditions at $\pm\infty$, the quantity
\eqn\Qdef{Q(\lambda )={\rm tr}\, U(-\infty ,\infty;\lambda )}
will be conserved
for any choice of the parameter $\lambda$. Indeed, $Q(\lambda )$ provides a
generating function for the conservation laws.

Unfortunately, once the field theory is restricted to a half line the same
argument cannot be used for the quantity
\eqn\halfpath{U(-\infty ,a;\lambda )={\rm P}\exp \int_{-\infty}^{a}\, a_1 dx^1\
.}
On the other hand, consider the path-ordered exponential around the
closed contour consisting of the following pieces $H_i:\ -\infty <x^1\le a;\
x^0=x^0_i, \ i=1,2$, $V_{-\infty}:\ x^1 =-\infty;\ x^0_1\le x^0\le x^0_2$
and $V_a:\ x^1=a;\ x^0_1\le x^0\le x^0_2$. Since the contour is closed, and
since
the gauge field $a_\mu$ has zero curvature inside the contour, the
path-ordered exponential around the contour is unity. That is, explicitly,
\eqn\contour{{\rm P}\exp\int_{V_{-\infty}}\, a_0 dx^0\ =\left({\rm P}
\exp\int_{H_1}
\, a_1 dx^1\right)
\, \left({\rm P}\exp\int_{V_{a}}\, a_0 dx^0\right)
\, \left({\rm P}\exp -\int_{H_2}\, a_1 dx^1\right).}
Choosing $\phi$ and its derivatives to vanish as $x^1\rightarrow -\infty$
guarantees the left hand side of \contour\ is unity. If in addition
the bondary condition could be used to show
that $a_0$ was a pure gauge at $x^1=a$, it would be possible to
write the middle term in the right hand side of \contour\ as a
product of group elements:
$${\rm P}\exp\int_{V_{a}}\, a_0 dx^0=G(x^0_1)G^{-1}(x^0_2).$$
Under these circumstances, eq\contour\ would imply
$$\left({\rm P}\exp \int_{H_2}\, a_1 dx^1\right)\, G(x^0_2)\,=\,
\left({\rm P}\exp\int_{V_{-\infty}}\, a_0 dx^0\right)\, G(x^0_1).$$
In other words, the combination
$$\left({\rm P}\exp \int_{-\infty}^{a}\, a_1 dx^1\right)\, G(x^0)$$
would be conserved. Unfortunately, the Lax pair is not in a suitable
form for the latter part of the argument since it does
not take into account the boundary condition at $x^1=a$. In fact,
although this
argument is appealing it requires modification to be useful.

To construct a modified Lax pair including the boundary condition,
it is first of all convenient to
consider an additional  special point $x^1=b\ (>a)$ and two overlapping
regions $R_-:\ x^1\le (a+b+\epsilon )/2;\ $ and $R_+:\ x^1\ge (a+b-\epsilon
)/2$.
The second region will be regarded as a reflection of the first,
in the sense that if $x^1\in R_+$, then
\eqn\reflectphi{\phi (x^1)\equiv\phi (a+b-x^1).}
The regions overlap in a small interval surrounding the midpoint of $[a,b]$.
Then, in the two regions define:
\eqn\newlax{\eqalign{&R_-:\qquad \widehat a_0=a_0 -{1\over 2}\theta (x^1-a)
\left(\partial_1\phi +
{\partial\B\over\partial\phi}\right)\cdot H \qquad
\widehat a_1=\theta (a-x^1)a_1\cr
&R_+:\qquad \widehat a_0=a_0 -{1\over 2}\theta (b-x^1)
\left(\partial_1\phi -
{\partial\B\over\partial\phi}\right)\cdot H \qquad
\widehat a_1=\theta (x^1-b)a_1.\cr}}
Then, it is clear that in the region $x^1<a$ the Lax pair \newlax\ is
the same as the old but, at $x^1=a$ the derivative of
the $\theta$ function in the zero curvature condition enforces the boundary
condition \boundary . Similar statements hold for $x^1\ge b$
except that the
boundary condition at $x^1=b$ is slightly different  in order to
accommodate the reflection condition \reflectphi .

On the other hand, for $x^1\in R_-$ and $x^1>a$, $\widehat a_1$ vanishes
and therefore the zero curvature condition merely implies that $\widehat a_0$
is independent of $x^1$. In turn, this fact implies that $\phi$ is
independent of $x^1$ in this region. Similar remarks apply to the region
$x^1\in R_+$ and $x^1<b$. Hence, taking into account the reflection principle
\reflectphi , $\phi$ is independent of $x^1$ throughout the interval $[a,b]$,
and equal to its value at $a$ or $b$. For general boundary conditions, a glance
at
\newlax\ reveals that the gauge potential $\widehat a_0$ is different in the
two
regions $R_\pm$. However, to maintain the zero curvature condition over the
whole
line the values of $\widehat a_0$ must be related by a gauge transformation
on the overlap. Since $\widehat a_0$ is in fact independent of $x^1\in [a,b]$
on both patches, albeit  with a different value on each patch,
the zero curvature condition effectively requires the existence of
a gauge transformation $\K$ with the property:
\eqn\Kdef{\partial_0 \K =\K\, \widehat a_0(x^0,b) -\widehat a_0(x^0,a)\, \K .}
The group element $\K$ lies in the group $G$ with Lie algebra $g$, the
Lie algebra whose roots define the  affine Toda theory.

Next, consider the analogue of \Qdef\ in the present context:
\eqn\Qhatdef{\widehat Q(\lambda ) ={\rm tr}\left( U(-\infty ,a;\lambda )\
\K \ U(b,\infty ;\lambda )\right).}
Making the usual assumptions concerning the fields at $x^1 =-\infty$ and using
\Kdef , $\widehat Q(\lambda )$ is time-independent. Moreover, the
reflection principle \reflectphi\ and conjugation properties \conj\
may be used to rewrite $U(b,\infty ;\lambda )$,
initially defined as a path-ordered exponential over $R_+$, as a
path-ordered exponential over $R_-$. Explicitly,
\eqn\upm{U(b,\infty ;\lambda )=\left({\rm P}\exp\int_{-\infty}^a
a_1^\dagger (\lambda ) dx^1\right)^\dagger\, =\, \left({\rm P}
\exp\int_{-\infty}^a
a_1 (1/\lambda ) dx^1\right)^\dagger .}
Hence, a more convenient expression for $\widehat Q(\lambda )$ is
\eqn\Qalt{\widehat Q(\lambda )={\rm tr}\left( U(-\infty ,a;\lambda )\
\K \ U^\dagger(-\infty, a ;1/\lambda )\right).}
This definition of  $\widehat Q(\lambda )$ is reminiscent of formulae
introduced previously by Sklyanin and Tarasov \refs{\rSl} for
other integrable models, but it is not the same.
These authors  appear to use the inverse for the second factor, not the
hermitian conjugate. However, this choice  would not be correct in the present
analysis. Indeed, the difference is crucial for Toda field theory.

In order to analyse further the restrictions imposed as a consequence of
requiring \Kdef , it is useful to make a couple of additional assumptions.
In particular, the gauge transformation $\K$ will be assumed to be
independent of both $x^0$ and the fields $\phi$ or their derivatives.
With these assumptions, and using the explicit expressions for
$\widehat a_0$, eq\Kdef\ reads,
\eqn\Kdefa{{1\over 2}\left[\K (\lambda ),\,
{\partial\B\over\partial\phi}\cdot H\right]_+=-\,\left[\K (\lambda )
,\, \sum_0^r
\sqrt{m_i}(\lambda E_{\alpha_i}-1/\lambda \ E_{-\alpha_i})
e^{\alpha_i\cdot\phi /2}\right]_-,}
where the field dependent quantities are evaluated at the boundary $x^1=a$.
Eq\Kdefa\ is rather stringent since the boundary term $\B$ does not
depend on the spectral parameter $\lambda$.
Clearly, one solution is always
\eqn\trivial{\K =1, \qquad \B = {\rm constant}, \qquad {\rm ie}\ \partial_1
\phi \Big|_a=0.}

To find other solutions, one might begin by noting that if $\K$ solves \Kdefa\
then so does $c\K$, where $c$ is any element of the centre of the group $G$.
Suppose $\K (0)$ exists, then \Kdefa\ implies
$$\left[\K (0),\, \sum_0^r
\sqrt{m_i}\, E_{-\alpha_i}\,
e^{\alpha_i\cdot\phi /2}\right]_-=\, 0,$$
which in turn implies $\K (0)$ is a central element of $G$ (remember,
$\K$ has been assumed
to be functionally independent of the fields $\phi$ and therefore
$\K (0)$ must commute with each of the generators $E_{-\alpha_i}$). In view of
the
ambiguity mentioned above, $\K (0)$ may be taken to be unity.

Next, suppose $\K$ has the form
\eqn\Kshape{\K (\lambda )=\exp \left(\sum_0^\infty\, k_n\lambda^n\right) ,}
substitute into \Kdefa , and solve order by order in $\lambda$. The $
\lambda^{-1}$ term is automatic, but order $\lambda^0$ yields
\eqn\orderzero{{\partial\B\over\partial\phi}\cdot H =\left[k_1,\,
\sum_0^r
\sqrt{m_i}\, E_{-\alpha_i}
e^{\alpha_i\cdot\phi /2}\right]_-\ ,}
requiring $k_1$ to have the form
\eqn\kone{k_1=\sum_0^r\, B_iE_{\alpha_i},}
and also implying
$${\partial\B\over\partial\phi}=\sum_0^r\, B_i\sqrt{n_i\over 2|\alpha_i|^2}\
\alpha_i\ e^{\alpha_i\cdot\phi /2}.$$
Clearly, the boundary term $\B$ is forced to have the form given in
\allboundary .
Using \orderzero , the order $\lambda^1$ terms are
$$\left[k_2,\,  \sum_0^r
\sqrt{m_i}\, E_{-\alpha_i}
e^{\alpha_i\cdot\phi /2}\right]_-\, =\, 0,$$
which implies (since $k_2$ is in the Lie algebra $g$), $k_2=0$. Actually, this
should have been expected on noting that if $\K (\lambda )$ solves \Kdefa ,
so does $\K^{-1}(-\lambda )$.

The terms of order $\lambda^2$ are more interesting and lead to further
constraints on the boundary coefficients $B_i$. Using the previous
results, \orderzero\ and $k_2=0$, they may be written:
\eqn\ordertwo{\left[ k_3,\, \sum_0^r\sqrt{m_i}\, E_{-\alpha_i}
e^{\alpha_i\cdot\phi /2}\right]_-={1\over 12}\, \left[k_1,\, \left[k_1,\,
{\partial\B\over\partial\phi}\cdot H\right]\right]_-+\left[k_1,\,
\sum_0^r
\sqrt{m_i}\, E_{\alpha_i}
e^{\alpha_i\cdot\phi /2}\right]_- .}
Now consider how this equation is graded with respect to the {\it principal}
grading of $g$. On positive roots, this grading is simply the length of the
root;
on negative roots it is the negative of this; on the Cartan subalgebra it is
zero. The two terms on the right of \ordertwo\ have grade $2\ {\rm mod}\ h$
and therefore this must also be the grade of the left hand side. In other
words, $k_3$ must have grade $3\ {\rm mod}\ h$. ( Typically $h>3$; but note,
there are  algebras, for example $a_1$
and $a_2$, which are special cases in having no generators of grade 3.)
Therefore, in general,
\eqn\kthree{k_3=\sum_{l(\beta )=3\ {\rm mod}\ h}\, C_\beta\, E_\beta .}
Using \kthree , \kone , the expression for the boundary term, and the
Lie algebra relations
\eqn\Liealgebra{\left[ E_\alpha ,\, E_\beta \right] \, =\,
\epsilon (\alpha ,\beta )E_{\alpha +\beta},}
the equation corresponding to matching the coefficients of
$e^{\alpha_i\cdot\phi /2}$ in eq\ordertwo\
may be written
\eqn\ordertwoa{\eqalign{\sum_{l(\beta )=3\ {\rm mod}\ h}\,C_\beta\, &\epsilon
(\beta ,-\alpha_i ) E_{\beta -\alpha_i}\, =\, -{1\over 12}\sum_{j,k\ne i}
B_iB_jB_k (\alpha_i\cdot\alpha_j )\epsilon (\alpha_k ,\alpha_j )
\sqrt{n_i\over 2|\alpha_i|^2}E_{\alpha_j +\alpha_k}\cr
& +\sum_j\, B_j \epsilon (\alpha_i,\alpha_j )
\sqrt{n_i\over 2|\alpha_i|^2}\left({1\over 12}B_i^2
\,(\alpha_i^2 -\alpha_i\cdot\alpha_j)-
{|\alpha_i|^2\over 2}\right)E_{\alpha_i+\alpha_j}.\cr}}
There are two cases to consider, the simply-laced $ade$ series and the rest.
\bigskip
\noindent {\it Simply-laced roots}
\bigskip
In this case, consider the generator $E_{\alpha_i+\alpha_j}$. Since no
level three root
in a simply-laced root system can have the form $\beta =2\alpha_i+\alpha_j$,
(the squared length of such a root would be either 10 or 6), then this
generator cannot appear in the sum on the left hand side of \ordertwoa . Since
it clearly does not appear in the first term on the right hand side, its
coefficient in the second term must vanish, implying restrictions on the
constants $B_j$. Indeed, $\alpha_j$ must be adjacent to $\alpha_i$ on
the (extended) Dynkin-Kac diagram. Therefore, for a particular choice of root,
$\alpha_i$,
the coefficients corresponding to its neighbours $\alpha_j$ satisfy
$${\rm  either}\qquad B_j=0\qquad {\rm or}\qquad B_j^2=4.$$
Taking each point of the Dynkin-Kac diagram in turn leads to the conclusion
that
$${\bf  either}\quad B_j=0\quad {\rm for\ all}\ j
\quad{\bf or}\quad B_j^2=4\quad {\rm for\ all}\ j,$$
which translates to precisely the conclusion reached previously, eq\gboundary ,
by considering specific charges of low spin.
Eq\ordertwoa\ otherwise determines the coefficients $C_\beta$ in terms of
the constants $B_j$ and Lie algebra data.

A complete solution to \Kdefa\ for the $a$ series will be given in an appendix
together with  a conjecture for $d_n,\ n>4$. The latter is based on solving
\Kdefa\ completely for $n=4,5,6,7$.
These complete solutions are essentially unique and their existence
places no further constraints on the boundary data.
\bigskip
\noindent{\it Non simply-laced roots}
\bigskip
Here, the story is slightly different and the restrictions
on the constants $B_j$ are less stringent. The point is that in these cases,
for which there are simple roots of different lengths, $2\alpha_i +\alpha_j$
is also a root provided $\alpha_i$ is a short root adjacent  on the
Dynkin-Kac diagram to a long root $\alpha_j$. Hence, if $i$ corresponds
to a short root, \ordertwoa\ has a term on the left hand side
which matches a term
on the right for which $j$ is a long neighbour of $i$; therefore, there is no
corresponding constraint equation involving the boundary constants. When
$\alpha_i$ and $\alpha_j$ are adjacent but have the same length, the
constraints
appear as they did for the simply-laced cases. The results obtained by
analysing the constraint equations on a case by case basis are reported
in an appendix.

\newsec{Integrability}

An important aspect of integrability is that the generating functionals
of the conserved charges are in involution. In other words,
given the canonical equal-time Poisson brackets
\eqn\poisson{\left\{\phi (x, x^0),\,\phi (y, x^0)\right\}=0\quad
\left\{\dot\phi (x, x^0),\,\dot\phi (y, x^0)\right\}=0\quad
\left\{\phi (x, x^0),\,\dot\phi (y, x^0)\right\}=\delta (x-y),}
the charge generating functionals \Qdef\ satisfy
\eqn\Qinvolute{\left\{Q(\lambda ),\, Q(\mu )\right\}=0,}
for any choices of $\lambda$ or $\mu$. As a consequence, the conserved
charges themselves are in involution. The crucial step in proving
\Qinvolute , relies on establishing the formula
\eqn\classr{\left\{ U(\lambda ),^\otimes \, U(\mu )\right\}=
\left[r(\lambda /\mu),\, U(\lambda )\otimes U(\mu )\right]_-,}
where $U(\lambda )$ represents the path-ordered exponential defined in
\pathexp . For the details of this, see for example
\NRF\rFTa\FTa\refs{\rFTa ,\rOTb ,\rOTd}.
In particular, for affine Toda field theory, the classical r-matrix has
the form \refs{\rOTd}
\eqn\classrdef{\eqalign{r(\lambda /\mu )=
{\mu^h+\lambda^h \over \mu^h-\lambda^h}
\sum_{i=1}^r H_i & \otimes H_i\cr
+{2\over \mu^h-\lambda^h}&\sum_{\alpha >0}\,{|\alpha|^2\over 2}
\left(\lambda^{l(\alpha )}\mu^{h-l(\alpha )}E_\alpha \otimes E_{-\alpha}
+\lambda^{h-l(\alpha )}\mu^{l(\alpha )}E_{-\alpha }\otimes E_{\alpha}\right)\
,\cr}}
where the sum is over all positive roots of $g$. Notice that
\eqn\rherm{r^\dagger (\lambda /\mu )=-r ( \mu /\lambda ),}
a property which will be used below. It is also useful to abbreviate $r$ by
writing
\eqn\classrabb{r (\lambda /\mu )=\sum_i r_i(\lambda /\mu )\, g_i
\otimes g_i^\dagger ,}
where $g_i$ ranges over the generators of the Lie algebra $g$.

A generating functional for conserved quantities on a half line has been
defined in eq\Qalt\ but it remains to be seen if the corresponding
charges are in involution. To investigate this requires an evaluation of  the
Poisson bracket
$$\left\{U(\lambda )\K (\lambda ) U^\dagger (1/\lambda ),^\otimes\,
U(\mu )\K (\mu ) U^\dagger (1/\mu )\right\}, $$
using \classr\ repeatedly, where $U(\lambda )$ is to be understood in the sense
of
\Qalt . One obtains a number of terms which are conveniently written:
\eqn\Qmess{\eqalign{\sum_i\biggl(
&r_i(\lambda /\mu )\, \bigl(g_iU_\lambda\K (\lambda )
U^\dagger_{1/\lambda}\otimes g^\dagger_iU_\mu\K(\mu )U^\dagger_{1/\mu}
-U_\lambda g_i\K(\lambda )U^\dagger_{1/\lambda}\otimes
U_\mu g^\dagger_i\K(\mu )U^\dagger_{1/\mu}\bigr) \cr
&+r_i(\lambda \mu )\, \bigl(g_iU_\lambda\K (\lambda )
U^\dagger_{1/\lambda}\otimes
U_\mu\K(\mu )U^\dagger_{1/\mu}g_i
-U_\lambda g_i\K(\lambda )U^\dagger_{1/\lambda}\otimes
U_\mu \K(\mu )g_iU^\dagger_{1/\mu}\bigr)\cr
&+r_i(1/\lambda \mu )\, \bigl(U_\lambda\K (\lambda ) U^\dagger_{1/\lambda}
g^\dagger_i\otimes
g^\dagger_iU_\mu\K(\mu )U^\dagger_{1/\mu}
-U_\lambda \K(\lambda )g^\dagger_iU^\dagger_{1/\lambda}\otimes
U_\mu g^\dagger_i\K(\mu )U^\dagger_{1/\mu}\bigr)\cr
&+r_i(\mu /\lambda )\, \bigl(U_\lambda\K (\lambda )
U^\dagger_{1/\lambda}g^\dagger_i\otimes U_\mu\K(\mu )U^\dagger_{1/\mu}g_i
-U_\lambda \K(\lambda )g^\dagger_iU^\dagger_{1/\lambda}\otimes
U_\mu \K(\mu )g_iU^\dagger_{1/\mu}\bigr)\biggr) .\cr}}
Once the trace is taken in each of the two spaces of the tensor product,
and taking into account \rherm , the sum of the first terms in each parenthetic
pair exactly cancel. The remaining terms do not automatically sum to zero.
However, if it is further assumed, closely following Sklyanin \refs{\rSl}, that
\eqn\rKeq{\left[ r( \lambda /\mu ),\, \K^{(1)}(\lambda )\K^{(2)}(\mu )
\right]_-=\K^{(1)}(\lambda )\widetilde r (\lambda\mu )\K^{(2)}(\mu )-
\K^{(2)}(\mu )\widetilde r (\lambda\mu )\K^{(1)}(\lambda ),}
then these four terms also cancel. Such an arrangement, not involving  $U$, is
not unreasonable given that $r$ and $\K$ are independent of the fields $\phi$.
In eq\rKeq , it was convenient to define
$$\K^{(1)}(\lambda )=\K( \lambda )\otimes 1\qquad\K^{(2)}(\mu )=1\otimes
\K (\mu ),$$
to facilitate writing the terms on the right hand side, and
\eqn\rtilde{\widetilde r (\lambda\mu )=\sum_i r_i (\lambda\mu )g_i\otimes g_i
.}
Notice that the second factor in the tensor product differs from the
corresponding factor in \classrabb\ in not being conjugated.
Notice also, that these manipulations work because of the presence of
$U^\dagger$ in \Qalt , rather than $U^{-1}$.

These considerations require not only the existence of $\K$ in the sense of
\Kdef , but also the same assumption as before, namely,
that $\K$ is independent of the fields $\phi$, and their
time derivatives. Otherwise, there would be extra terms in \Qmess , since
one would have to worry about the Poisson brackets of $\K$ and the
other factors in $\widehat Q$. On the other hand, it was noted
previously that
given \Kdefa , the quantity $\K$ is essentially unique and the boundary
conditions on the fields are strongly restricted. There is a danger that
eq\rKeq\ is not compatible with these assumptions. Since \rKeq\ is bilinear
as far as the two $\K$'s are concerned, multiplying $\K$ by a central element
of
the group G (the only ambiguity in \Kdef ), has no effect whatsoever. It
therefore needs to be checked that \rKeq\ places no stronger constraints on the
boundary conditions or, otherwise fails to be compatible with $\K$.

In fact, as will be shown below,
the solutions already found for $\K$ satisfy \rKeq\ identically.

In a sense, \Kdefa\ is a more fundamental equation than the defining equation
\classr .  The argument could be turned around:
given $\K$, eq\rKeq\ may be regarded as an equation
for the classical r-matrix itself. This is reminiscent of an observation
made in \refs{\rCDRa} noting that once the reflection factors are known in
the quantum field theory then the bootstrap relations would imply the S-matrix
elements. From this point of view,
the reflection factors are  fundamental quantities.
\bigskip
\noindent{\it $\K$ - $r$ compatibility}
\bigskip

First, rewrite \Kdefa , taking into account the form of the boundary
condition and that $\K$ is independent of the fields, to
yield (for $i=0,\dots ,r$)
\eqn\Kdefb{\K(\lambda )\left(B_i{\alpha_i\cdot H\over |\alpha_i|^2}+
\lambda E_{\alpha_i}-{1\over\lambda}E_{-\alpha_i}\right)=
\left(-B_i{\alpha_i\cdot H\over |\alpha_i|^2}+
\lambda E_{\alpha_i}-{1\over\lambda}E_{-\alpha_i}\right)\K(\lambda ).}
Define the Lie algebra elements
\eqn\Xpm{
X_i^{\pm}(\lambda)=\pm \lambda\, B_i{\alpha_i \cdot H\over |\alpha_i|^2}
+\lambda^2 E_{\alpha_i}-E_{-\alpha_i}.}
In terms of these, \Kdefb\ reads
\eqn\Kdefc{K(\lambda ) X_i^+(\lambda ) K^{-1}(\lambda ) = X_i^-(\lambda ).}

First, it can be shown that the solution to \Kdefc\  is unique up to
multiplication by a central element of $G$. To demonstrate this, note first
that
$X_i^\pm(\lambda)$ can be used to generate
all of the Lie algebra $g$ by repeated commutation. In other words, if one
defines
$$X^{\pm}_{ij....k}=[X^{\pm}_i,[X^{\pm}_j,[.......,X^{\pm}_k]]..]$$
then there exists a subset ${\bar X}^{\pm}_i$ of all the $X^{\pm}_{ij...k}$
which span $g$. Clearly this is true at $\lambda=0$, since
$X^-_i(0)=-E_{-\alpha_i}$. The generators corresponding to the
negative simple roots may be used to
manufacture the generators for all the negative roots,
and then the highest weight
generator $E_{-\alpha_0}$ can be used with the negative
simple roots to obtain all the
positive roots and the Cartan subalgebra. This defines a spanning basis
${\bar X}^{\pm}_i(0)$. Consider the set ${\bar X}^{\pm}_i(\lambda)$.
In terms of an orthonormal basis $T_i$ for $g$,
$$
{\bar X}^{\pm}_i(\lambda)=A_{ij}(\lambda)T_j
$$
where $A_{ij}(\lambda)$ is a matrix with polynomial entries in $\lambda$.
The quantities
${\bar X}^{\pm}_i(\lambda)$ will span $g$ if and only if the matrix $A_{ij}$
is invertible. However, the determinant of $A_{ij}$ does not identically vanish
since it is non-zero at $\lambda=0$, so it must be a non-vanishing polynomial
in $\lambda$. Hence, except for a finite number of values of $\lambda$,
${\bar X}^{\pm}_i(\lambda)$ spans $g$.

Next, suppose there are two solutions $K_1,K_2$ to \Kdefc\ . Then
$$
(K_2^{-1} K_1)^{-1} X_i^+(\lambda)K_2^{-1} K_1=K_1^{-1} X_i^-(\lambda)
K_1=X_i^+(\lambda)
$$
and the same equation holds for $X^+_{ij....k}$, and in particular for
${\bar X}^+_i$, and so $g$ is fixed under conjugation by $K_2^{-1} K_1$.
It follows from Schur's lemma that $K_2^{-1} K_1$ must be in the centre
of $G$.

In order to demonstrate that  \Kdefc\ implies \rKeq\ it is convenient to
consider
$$\eqalign{L=&(K^{(1)}(\lambda) K^{(2)}(\mu))^{-1}r({\lambda /\mu})
K^{(1)}(\lambda) K^{(2)}(\mu)-r({\lambda /\mu})\cr
&\ \ +(K^{(1)}(\lambda))^{-1}{\tilde r}(\lambda\mu)K^{(1)}(\lambda)
-(K^{(2)}(\mu))^{-1}{\tilde r}(\lambda\mu)K^{(2)}(\mu),\cr}$$
and argue that $L$ vanishes.

As a first step, it may be shown using the explicit epression for
the classical $r$-matrix, eq\classrdef , that
\eqn\LX{ [L,X^-_i(\lambda)\otimes {\bf 1}+ {\bf 1}\otimes X^-_i(\mu)]=0.}

For the next step,  consider equation \LX\  at $\lambda=\mu$.
Along this line $L$ commutes with all generators of the form
$T_i\otimes{\bf 1}+{\bf 1}\otimes T_i$. This implies that it must be
proportional to the Casimir-like operator
${\bf C}=\sum_i T_i\otimes T_i$; that is
$$
L(\lambda,\lambda)=\gamma(\lambda)\sum_i T_i\otimes T_i
$$
To determine $\gamma(\lambda)$,
multiply this equation by $X^-_i(\lambda)\otimes X^-_j(\lambda)$
 and take the trace in both first and second space. The left hand side reads
$$\eqalign{
{\rm Tr}_{1,2}\left( X^-_i(\lambda)\otimes X^-_j(\lambda)\, L(\lambda,\lambda)
\right)&=\
{\rm lim}_{\lambda\to\mu}{\rm Tr}_{1,2}\left( X^-_i(\lambda)\otimes X^-_j(\mu)
L(\lambda,\mu)\right)\cr
&=\  {\rm lim}_{\lambda\to\mu}{\rm Tr}_{1,2}\left(X^+_i(\lambda)\otimes
X^+_j(\mu)
r({\lambda /\mu})\right)\cr
&\ \ -{\rm lim}_{\lambda\to\mu}{\rm Tr}_{1,2}\left(X^-_i(\lambda)\otimes
X^-_j(\mu)r({\lambda /\mu})\right)\cr
&\ \ +{\rm lim}_{\lambda\to\mu}{\rm Tr}_{1,2}\left(X^+_i(\lambda)\otimes
X^-_j(\mu){\tilde r}({\lambda\mu})\right)\cr
&\ \ -{\rm lim}_{\lambda\to\mu}{\rm Tr}_{1,2}\left(X^-_i(\lambda)\otimes
X^+_j(\mu){\tilde r}({\lambda\mu})\right)\cr}
$$
but, using the cyclic property of trace
and explicitly substituting in $r({\lambda /\mu})$ and
${\tilde r}(\lambda\mu)$ the first two terms and last two terms cancel.
Therefore,
the left hand side vanishes.
If $B_k=0$ then choosing $i=j=k$ gives a non-vanishing coefficient to
$\gamma(\lambda)$, whilst if all the $B_i$ are non-vanishing it can still
be arranged for the coefficient to be non-vanishing by choosing $i$ and $j$ to
correspond to neighbouring points on the Dynkin diagram. Thus,
$L(\lambda,\lambda)$ vanishes.

Finally, the proof may be completed by demonstrating that $L(\lambda,\mu)$ also
vanishes away from the line $\lambda=\mu$. To do this, consider \LX\
as an equation for the variable $L$. The solutions form a vector space with
at most dimension one. Suppose there are two linearly
independent solutions $S_1(\lambda,\mu),\ S_2(\lambda,\mu) \in g\otimes g$.
A linear combination of these, $S$, may always be found such that
${\rm Tr}_{1,2}( S {\bf C})=0$; for, if
$$\eqalign{
{\rm Tr}_{1,2}( S_1 {\bf C})&= s_1(\lambda,\mu)\cr
{\rm Tr}_{1,2}( S_2 {\bf C})&= s_2(\lambda,\mu)\cr}
$$
where $s_1,s_2$ are non-vanishing, simply take $S=s_2 S_1- s_1 S_2$, while
if $s_1$ vanishes, take $S=S_1$ (or, if  $s_2$ vanishes, take $S=S_2$).
Now working in
some matrix representation of $g\otimes g$, the entries of $S$ are
rational functions of $\lambda$ and $\mu$, since $S$ is the solution to a
set of simultaneous linear equations. So by multiplying by suitable powers
of $(\lambda-\mu)$,  $S$ can be arranged to be finite and non-zero at
generic points on the line $\lambda=\mu$. Along the line $\lambda=\mu$,
$S$ should be proportional to ${\bf C}$. However,
since ${\rm Tr}_{1,2}( S {\bf C})=0$,
it would follow that $S$ vanishes, which is a contradiction.
If the space of solutions to \LX\ is zero-dimensional away from
the line $\lambda=\mu$ it follows that $L$ must vanish away from that line
and the result is proved. If it is one dimensional,
denote a basis vector ${\bf C}(\lambda,\mu)$ and by
the above, it can be normalised so that it coincides with ${\bf C}$ for
$\lambda=\mu$. Therefore
$$
L=\gamma(\lambda,\mu){\bf C}(\lambda,\mu).
$$
Proceeding similarly, as for the case $\lambda=\mu$, multiply
this equation by $X^-_i(\lambda)\otimes X^-_j(\mu)$ and take the trace in
both spaces. The left hand side vanishes, while the
coefficient of $\gamma(\lambda,\mu)$ is a continous function of $\lambda$
and $\mu$. For appropriate $i$ and $j$, it is non-vanishing along the
line $\lambda=\mu$ and, by continuity non-vanishing in some neighbourhood
of this line. It follows that $L$ must vanish for generic $\lambda$ and $\mu$.

\newsec{Summary and discussion}

In this article much of the detail omitted from \refs{\rCDRa} has
been included. However, the approach advocated there is quite limited since
the possibility remains that new conditions might be needed to guarantee the
existence of charges of spin greater than four. To compensate for this
limitation, a development of the Lax pair idea has been presented which is
able to take into account the boundary condition. Within this scheme, in
order to make  progress, it is nevertheless necessary to make some
natural assumptions which lead eventually to the basic equations \Kdefa\
or  \Kdefb . Given these assumptions, the same restrictions on the boundary
conditions are obtained in those situations where both approaches are feasible.
However, the Lax pair scheme also allows conjectures to be formulated in all
other cases.

The set of equations \Kdefb\ are interesting by themselves. Although a number
of exact solutions have been found to these
equations, for any given Lie algebra the solutions have very little freedom,
and indeed there is not yet a proof of existence in the general case. It
is intriguing that the solutions to \Kdefb\ are also compatible with \rKeq\
since this was not guaranteed by the method for constructing $\K$. The
relationship
between the two quantities $r$ and $\K$, and the defining
equations for $\K$,  needs to be clarified.

All the considerations of this paper have been entirely classical and a
question remains concerning the compatibility of the general boundary
conditions
and quantum integrability. This is a difficult question to which we hope to
return in the future.

\bigskip

\noindent{\bf Acknowledgements}
\bigskip
One of us (RHR) wishes to thank the UK Engineering and Physical Sciences
Research Council
for a Research Assistantship.

\appendix{A}{}

In this Appendix it is explained how eqs \abcda\ are solved to find
expressions for
$T_{\pm5}$ for $a_{n}^{(1)}$, $d_{5}^{(1)}$ and $e_{6}^{(1)}$.

It is
convenient to take $B_{(abcd)}$ to
vanish; a totally symmetric part of $B_{abcd}$ can always be added if it makes
the final expressions that we find more compact. After introducing a notation
analogous to eq \abdef\  eqs \abcda\ imply
\eqn\appeen{\eqalign{D_{ijk} = & 4  E_{k(i} C_{j)k}, \cr
B_{ijkl} - B_{(jk)li} = & - D_{li(j} C_{k)l} +
{1 \over 2} C_{l(j} D_{k)li} + {1 \over 2} D_{l(jk)} C_{il} \cr
& \ \ \ \ \ \ \ \ \ \ \ \ \ \ \ \ + 2  E_{li} C_{lj} C_{lk}
- 2 \beta^2 E_{l(j} C_{k)l} C_{il}, \cr
A_{ijklm} = & - {1 \over 3} B_{i(jkl} C_{m)i}
+ {1 \over 3} C_{i(j} B_{klm)i}
+ {1 \over 3} D_{i(jk} C_l^{\; i} C_{m)}^{\;\; i}
- {2  \over 3} E_{i(j} C_{k}^{\; i} C_{l}^{\; i} C_{m)}^{\;\; i}. \cr }}
Here no summation over repeated indices is intended.
Symmetrising the second equation in $j$, $k$ and $l$ and using the fact that
$B_{ijkl}$ is symmetric in its last three indices and that its totally
symmetric part vanishes, it is found that
\eqn\apptwee{\eqalign{
B_{ijkl}  = & {1 \over 4}
\Bigl(
- D_{li(j} C_{k)l} +
{1 \over 2} C_{l(j} D_{k)li} + {1 \over 2} D_{l(jk)} C_{il} +
2  E_{li} C_{lj} C_{lk} +
\cr
& \ \ \ \ \ \ \ \ \ \ \ \ \ \ \ \ \ \ \
- 2  E_{l(j} C_{k)l} C_{il}
+ (l \rightarrow j \rightarrow k \rightarrow l)
+ (l \rightarrow k \rightarrow j \rightarrow l)
\Bigr). \cr}}
Hence, $D_{ijk}$, $B_{ijkl}$ and $A_{ijklm}$ can be calculated once
 $E_{ij}$ has been found. Before calculating $E_{ij}$ there are  a few
consistency
conditions which follow from eqs \appeen. Using
\eqn\al{\sum_{i=0}^{n} n_{i} \alpha_{i}^{a} = 0,}
one finds
\eqn\appdrie{\eqalign{0 & = \sum_{k=0}^{n} n_k D_{ijk} =
\sum_{k=0}^{n} 4 n_k  E_{k(i} C_{j)k}, \cr
0 & =
\sum_{l=0}^{n} n_l \left( B_{ijkl} - B_{(jk)li} \right)  \cr
& = \sum_{l=0}^{n} n_l \Bigl( -  D_{li(j} C_{k)l} +
{1\over 2} C_{l(j} D_{k)li} + {1\over 2} D_{l(jk)} C_{il} +
2  E_{li} C_{lj} C_{lk}
- 2  E_{l(j} C_{k)l} C_{il} \Bigr), \cr
0 & = \sum_{i=0}^{n} n_i A_{ijklm}  \cr
& = \sum_{i=0}^{n} n_i \Bigl( - {1 \over 3} B_{i(jkl} C_{m)i}
+ {1 \over 3} C_{i(j} B_{klm)i}
+ {1 \over 3} D_{i(jk} C_l^{\; i} C_{m)}^{\;\; i}
- {2  \over 3} E_{i(j} C_{k}^{\; i} C_{l}^{\; i} C_{m)}^{\;\; i}
\Bigr). \cr }}
Only relations which are obtained by summing over an index appearing twice
on the right hand side in eqs \appeen\ will be non-trivial. There are more
relations that have to be satisfied. From eq \appeen\ it is clear that
$D_{ijk}$ will be automatically symmetric in its first two indices. Eq
\apptwee\ will give a $B_{ijkl}$ which is automatically symmetric in its last
three indices, but the extra condition
\eqn\appvier{B_{(ijkl)}=0,}
has to be added separately since this is not automatically implied by eq
\apptwee . Finally, it is not clear
from the last of eq \appeen\ that $A_{ijklm}$ is symmetric in all its indices.
Hence, this symmetry must be imposed explicitly,
\eqn\appvijf{A_{ijklm}=A_{(ijklm)}.}
{}From the first of eqs \appdrie\ it follows that for $a_{n}^{(1)}$
the elements of
$E$ satisfy $E_{ij}=E_{i+1 \; j+1}$ for $i,j=0, \ldots ,n$. For $d_{5}^{(1)}$,
on the other hand, (with 4,5 labelling the simple roots on the fork of the
Dynkin diagram)
 $E$ has to be proportional to
$$
E \sim \pmatrix
{0 & 0 & 0 & 0 & +1 & -1 \cr
0 & 0 & 0 & 0 & -1 & +1 \cr
0 & 0 & 0 & 0 & 0 & 0 \cr
0 & 0 & 0 & 0 & 0 & 0 \cr
-1 & +1 & 0 & 0 & 0 & 0 \cr
+1 & -1 & 0 & 0 & 0 & 0 \cr},
$$
while for $e_{6}^{(1)}$, (with 1,4 and 2,5 labelling  simple roots
corresponding
to spots
on the long branches of the Dynkin diagram, working towards the centre, and
6 being the centre spot),
$$
E = \pmatrix
{0 & -2(a+b) & +2(a+b) & 0 & -a & +a & 0 \cr
+2(a+b) & 0 & -2(a+b) & +a & 0 & -a & 0 \cr
-2(a+b) & +2(a+b) & 0 & -a & +a & 0 & 0 \cr
0 & -a & +a & 0 & -b & +b & 0 \cr
+a & 0 & -a & +b & 0 & -b & 0 \cr
-a & +a & 0 & -b & +b & 0 & 0 \cr
0 & 0 & 0 & 0 & 0 & 0 & 0 \cr},
$$
with $a$ and $b$ arbitrary constants. From these expressions for $E$, $D$
can be calculated from the first of eq \appeen . Substituting $D$ and
$E$ in the
second of eq \appdrie\ restricts $E$, and therefore also $D$, even more in the
case of $a_{n}^{(1)}$ and $e_{6}^{(1)}$. For $a_{n}^{(1)}$,
$$
E \sim \pmatrix
{0 & 2(n-3) & 4(n-3) & 4(n-5) & \ldots & \ldots \cr
 & & & & & \cr
-2(n-3) & 0 & 2(n-3) & 4(n-3) & \ldots & \ldots \cr
 & & & & & \cr
-4(n-3) & -2(n-3) & 0 & 2(n-3) & \ldots & \ldots \cr
 & & & & & \cr
-4(n-5) & -4(n-3) & -2(n-3) & 0 & \ldots & \ldots \cr
 & & & & & \cr
-4(n-7) & -4(n-5) & -4(n-3) & -2(n-3) & \ldots & \ldots \cr
\vdots & -4(n-7) & -4(n-5) & -4(n-3) & \ldots & \ldots \cr
4(n-7) & \vdots & -4(n-7) & -4(n-5) & \ldots & \ldots \cr
4(n-5) & 4(n-7) & \vdots & -4(n-7) & \ldots & \ldots \cr
4(n-3) & 4(n-5) & 4(n-7) & \vdots & \ldots & \ldots \cr
 & & & & & \cr
2(n-3) & 4(n-3) & 4(n-5) & 4(n-7) & \ldots & \ldots \cr}.
$$
Hence  $a_{3}^{(1)}$ has no spin four charge, as was to be expected.
For $e_{6}^{(1)}$, $b=0$ and, up to one overall constant $E$ has now
been uniquely determined. The quantities $B$ and $A$ may now
be calculated from \appeen\ and
\apptwee\ and eqs\appvier\ and \appvijf\ checked.
Hence for $a_{n}^{(1)}$, $d_{5}^{(1)}$ and $e_{6}^{(1)}$ there is exactly
exactly one current.

At first sight, it appears surprising that this overdetermined set of
equations does give a
solution in the cases where a spin four charge is expected. This is, however,
less remarkable than it seems. Eqs \abcda, from which eqs\appeen\ were found,
are
the conditions under which $\partial_{\mp} T_{\pm5} = \d \Theta_{\pm 3}$ for
some
$\Theta_{\pm3}$. Calculating the left hand side of this equation by
substituting the Ansatz for $T_{\pm5}$ as given in \Tfive\ yields a total
derivative plus terms proportional to $(\d \phi)^4$, $(\d \phi)^2 \d^2 \phi$
or $(\d^2 \phi)^2$. All these terms have to vanish and this gives
the three equations \abcda. However, these three types of terms can be
transformed into each other by adding total derivatives. This means that terms
will shift from one equation to another, which in turn introduces arbitrary
parameters. The details are omitted for lack of space,
but one  needs each of the eqs\appeen, \apptwee,
\appdrie, \appvier\ and \appvijf\ to determine first these arbitrary parameters
and then $E$, $D$, $B$ and $A$. Only for $a_{n}^{(1)}$, $d_{5}^{(1)}$ and
$e_{6}^{(1)}$ is there a non-trivial solution.

Once $E_{ij}$, $D_{ijk}$, $B_{ijkl}$ and $A_{ijklm}$ are found,
$E_{ab}$, $D_{abc}$, $B_{abcd}$ and $A_{abcde}$ are
obtained by inverting the transformation
\abdef
\eqn\abdefinv{E_{ab}=\sum_{i,j=1}^{n}
\left( \alpha^{-1} \right)_{a}^{\; i}
\left( \alpha^{-1} \right)_{b}^{\; j} E_{ij},}
etc. This will lead to expressions for the current $T_{\pm5}$.
For $a_{n}^{(1)}$
this expression is given in \Tfivean. A totally symmetric term has been
added to
$B_{abcd}$ in that case to make the expression more appealing. The same was
done
in the case of $d_{5}^{(1)}$ leading to
\eqn\Tfivedvijf{\eqalign{
T_{\pm 5} =
& - {1 \over 4 \sqrt{2}}
\left\{
\left[ \d \phi_1 + \d \phi_3 \right]^2 -2 \left( \d \phi_2 \right)^2
\right\}
\left[ \d \phi_1 - \d \phi_3 \right]
\left[
\left( \d \phi_s \right)^2 - \left( \d \phi_{\bar{s}} \right)^2
\right]
\cr
& + \left[
\d^2 \phi_s \d \phi_s - \d^2 \phi_{\bar{s}} \d \phi_{\bar{s}}
\right]
\left[ \d \phi_1 - \d \phi_3 \right] \d \phi_2
\cr
& + {1 \over 2 \sqrt{2}}
\left\{
\left[ \d^2 \phi_1 + \d^2 \phi_3 \right] \d \phi_2
- \d^2 \phi_2 \left[ \d \phi_1 + \d \phi_3 \right]
\right\}
\left[ \left( \d \phi_s \right)^2 - \left( \d \phi_{\bar{s}} \right)^2 \right]
\cr
& - {1 \over 4}
\left[ \d^2 \phi_{\bar{s}} \d \phi_s - \d^2 \phi_s \d \phi_{\bar{s}} \right]
\left[
\left[ \d \phi_1 + \d \phi_3 \right]^2 -
2 \left( \d \phi_2 \right)^2
\right]
\cr
& - {1 \over \sqrt{2}}
\left[ \d^2 \phi_{\bar{s}} \d \phi_s - \d^2 \phi_s \d \phi_{\bar{s}} \right]
\left[ \left( \d \phi_1 \right)^2 - \left( \d \phi_3 \right)^2 \right]
\cr
& + \left[
\left( \d^2 \phi_s \right)^2 - \left( \d^2 \phi_{\bar{s}} \right)^2
\right]
\left\{
\left[ \d \phi_1 + \d \phi_3 \right]
+ {1 \over \sqrt{2}} \left[ \d \phi_1 - \d \phi_3 \right]
\right\} \cr
& + \left[ \d^2 \phi_1 + \d^2 \phi_3 \right]
\left[ \d^2 \phi_s \d \phi_s - \d^2 \phi_{\bar{s}} \d \phi_{\bar{s}} \right]
- \sqrt{2} \d^2 \phi_2
\left[ \d^2 \phi_{\bar{s}} \d \phi_s - \d^2 \phi_s \d \phi_{\bar{s}} \right]
\cr
& + \d^4 \phi_s \d \phi_{\bar{s}} - \d^4 \phi_{\bar{s}} \d \phi_s .\cr}}
For $e_{6}^{(1)}$ the expression is far too long to be reported here.

\appendix{B}{Explicit expressions for $\K$ in special cases}

It is possible to solve \Kdefb\ exactly for the special cases $a_n^{(1)}$
and $d_{4,5,6,7}^{(1)}$, and to conjecture a solution for $d_n^{(1)}$ on the
basis of
the latter.

It is convenient to work within the smallest representation of
$a_n$ ($n+1$-dimensional) and to use a representation for
the generators corresponding to the simple roots of the form
$$\left(E_{\alpha_i}\right)_{jk}=\delta_{j\ i-1}\delta_{k\ i}\qquad i,j,k=
0,1,2 \dots ,n \ {\rm mod}\ h=n+1.$$
It is also convenient to set $B_i=2C_i$,  to define
$C=\prod_0^r\, C_i^{n_i}$, to let $l(\alpha )$ denote the level
of a root, and to define $l_i(\alpha )$ to be the integer coefficient
of the root $\alpha_i$ in the simple root decomposition of $\alpha$.
Then,
the solution for $\K$ satisfying
$$\K^\dagger (\lambda )=\K( 1/\lambda ),$$
is
\eqn\Kan{1\, +\, \sum_{\alpha >0}\, \prod_i C_i^{l_i(\alpha )}\, \left[\left(
{2\over 1+C\lambda^h}\right) (-\lambda )^{l(\alpha )}
E_\alpha +\left({2\over 1+C\lambda^{-h}}\right) (-1/\lambda )^{l(\alpha )}
E_{-\alpha}\right].}
This case is probably the simplest, in the sense that $\K$
is expressible in terms of Lie algebra generators alone. For that reason, the
formula has been established by direct calculation for every $n$.

For the case of $d_n$, the situation is somewhat more complicated. There,
it is convenient to choose to work in the $2n$-dimensional representation
and to set
$$\eqalign{&\left(E_{\alpha_i}\right)_{jk}=\delta_{j\ i}\delta_{k\ i+1}+
\delta_{j\ n+i+1}\delta_{k\ n+i }\qquad i=1,2,\dots ,n-1
\qquad j,k=1,2,\dots ,2n\cr
&\left(E_{\alpha_0}\right)_{jk}=\delta_{j\ n+2}\delta_{k\ 1}+
\delta_{j\ n+1}\delta_{k\ 2 }\cr
&\left(E_{\alpha_n}\right)_{jk}=\delta_{j\ n-1}\delta_{k\ 2n}+
\delta_{j\ n}\delta_{k\ 2n-1 },\cr} $$
corresponding to the choice of roots:
$$\alpha_i=e_{i}-e_{i+1}\qquad i=1,2,\dots , n-1,\qquad
\alpha_0=-(e_1+e_2), \quad
\alpha_n=e_{n-1}+e_n,$$
where $e_i,\ i=1,2, \dots ,n$ are orthonormal vectors.

Then, for
$d_{\rm even}$, $\K$ is
conjectured to have the form:
\eqn\Kdeven{\eqalign{1\,  +\, \sum_{\alpha >0}\, \prod_i C_i^{l_i(\alpha )}\,
&\left[
\left(
{2\over 1+C\lambda^h}\right) (-\lambda )^{l(\alpha )}
\, \widehat{E}_\alpha +\left({2\over 1+C\lambda^{-h}}\right)
(-1/\lambda )^{l(\alpha )}
\, \widehat{E}_{-\alpha}\right]\cr
&+ \, {4C_nC_{n-1}\over (1+C\lambda^h)(1+C\lambda^{-h})}
\, \sum_{\beta }\,
 \lambda^{\widehat{l}(\beta  )}
C^{l(\beta  )/2}\,  \widehat{E}_{\beta },\cr}}
where the sum over $\beta$ in the last term refers to the set of
vectors $\pm 2e_i$, which are not roots, but expressible in terms
of the roots:
$$\eqalign{2e_n=\alpha_n-\alpha_{n-1},\ 2e_{n-1}=\alpha_{n-1}+\alpha_n,\
&2e_{n-2}=2\alpha_{n-2}+\alpha_{n-1}+\alpha_n, \dots\cr
&\dots , 2e_1=2\alpha_1+\dots +
2\alpha_{n-2}+\alpha_{n-1}+\alpha_n,\cr}$$
and the quantity $\widehat{l}(\beta )$ is defined by
$$\widehat{l}(\beta )=\cases{l(\beta ) &if $l(\beta )=0$\ mod 4\cr
                             h-l(\beta )&if $l(\beta )=2$\ mod 4.\cr}$$
The matrices
corresponding to these vectors are not Lie algebra generators but have
the form
$$\left(\widehat{E}_{2e_i}\right)_{jk}=\delta_{j\ i}
\delta_{k\ n+i}=\left(\widehat{E}_{-2e_i}\right)_{kj}
\qquad i=1,\dots ,n,\quad j,k=1,\dots , 2n.$$
Finally,
the other matrices $\widehat{E}_\alpha$ are either generators, or conjugate to
generators; in fact, they are given by
$$ \widehat{E}_\alpha =\Omega^{1+l(\alpha )}\, E_\alpha\, \Omega^{-1-l(\alpha
)},$$
where
$$\Omega ={\rm diag}(1,1,1,\dots ,1,-1,1,-1,-1,-1,\dots ,-1).$$
However, why this should be the case remains a mystery. The expression
has been checked for the cases $d_4, d_6$  using Mathematica.

For $d_4$, the
following relation also holds
$$\K (\lambda )\K (-\lambda )=\left({1+C\lambda^6\over 1-C\lambda^6}
\right)^2\, 1.$$
An equation of this form was to be expected, replacing
$\lambda\rightarrow -\lambda$ in
\Kdefb , given the uniqueness of the solution up to a scalar factor
but the corresponding relation for all the other cases has not
been established.

For $d_{\rm odd}$, the conjectured solution is a similar expression to
\Kdeven , except for the
extra $2n$ terms which have a different, simpler form:
\eqn\Kdodd{\eqalign{1\,  +\, \sum_{\alpha >0}\, \prod_i C_i^{l_i(\alpha )}\,
&\left[
\left(
{2\over 1+C\lambda^h}\right) (-\lambda )^{l(\alpha )}
\, \widehat{E}_\alpha +\left({2\over 1+C\lambda^{-h}}\right)
(-1/\lambda )^{l(\alpha )}
\, \widehat{E}_{-\alpha}\right]\cr
&+C_{n}C_{n-1}\sum_{l(\beta )=2\ {\rm mod}\ 4}\left(\,  {\lambda^{l(\beta
)}\over 1+C\lambda^h}\widehat{E}_{\beta }+{\lambda^{-l(\beta )}\over
1+C\lambda^{-h}}\widehat{E}_{-\beta }\,\right).\cr}}
This has been checked explicitly for $d_5$ and $d_7$.

\appendix{C}{The special case $a_{2}^{(2)}$}

It is quite instructive to consider the spin five charges for $a_2^{(2)}$
explicitly.

To facilitate calculation, it is convenient to choose a normalisation
for the roots
for which the equation of motion is
\eqn\bd{\partial_+\partial_-\phi = -{1\over 2}V^\prime(\phi ), \qquad
V(\phi )=e^{2\phi}+2e^{-\phi}.}
Then, the appropriate spin $\pm 6$ densities are
\eqn\tsix{T_{\pm 6}=(\partial_\pm\phi )^6 - 5(\partial_\pm\phi )^3
\partial^3_\pm\phi
+5 (\partial^2_\pm\phi )^3 +3 (\partial^3_\pm\phi )^2,}
satisfying
\eqn\tsixtheta{\partial_\mp T_{\pm 6}=\partial_\pm \Theta_{\pm 4}}
where
$$\Theta_{\pm 4}=-{1\over 8}\left[ 4 (\partial_\pm\phi )^2 \partial^2_\pm\phi
(-15 V^\prime +6 V^{\prime\prime\prime}) +12 (\partial^2_\pm\phi )^2
V^{\prime\prime}
+ (\partial_\pm\phi )^4 (10V^{\prime\prime}-6
V^{\prime\prime\prime\prime})\right] .$$
Insisting the combination $T_6-T_{-6}+\Theta_4-\Theta_{-4}$ is a total time
derivative
in the presence of a boundary term at $x^1=0$ requires the boundary term  in
the
lagrangian to have the form
\eqn\atwotwo{{\cal B}=A_1e^\phi +A_0e^{-\phi /2}}
where
$$A_0(A_1^2-2)=0.$$

The above normalisation is less convenient for solving \Kdefa , however.

In order to discover an expression for $\K$ in this case, it is first
necessary to obtain the data for \Kdefb\ by folding $a_2^{(1)}$. Ie, if
$\alpha_0,\alpha_1$ and $\alpha_2$ are the simple roots of $a_2^{(1)}$,
the relevant roots for $a_2^{(2)}$ are $\beta_0=(\alpha_0+\alpha_2)/2$
and $\beta_1=\alpha_1$. It is convenient to take the corresponding
generators to be
$$E_{\beta_0}=\sqrt{2}(E_{\alpha_0}+E_{\alpha_2})$$
and to work in the three-dimensional representation of $a_2$, where
$$\beta_1\cdot H={\rm diag}(1,-1,0)\quad{\rm and}\quad
\beta_0\cdot H=-{1\over 2}{\rm diag}(1,-1,0),$$
and to set $B_0=-\sqrt{2}C_0,\ B_1=2C_1$. Then, \Kdefb\ has a solution,
unique up to a scalar factor, provided
$$C_0(C_1^2-1)=0,$$
which is the same condition as the above once the differing normalisations
are accounted for.
The solution for $\K$ is:
$$\eqalign{(C_1&-\lambda^3){\bf 1}\ +\cr
&\pmatrix{0&
{-2\lambda (C_1^2-C_1\lambda^3(1-C_0^2)+C_0^2\lambda^6)\over 1+\lambda^6}
&-2C_0\lambda^2\cr
{2\lambda^2(C_0^2C_1+(C_0^2-C_1^2)\lambda^3+C_1\lambda^6)\over
1+\lambda^6}&0&2C_0C_1\lambda\cr
2C_0C_1\lambda&-2C_0\lambda^2&
{-2\lambda^3(C_0^2-C_1^2 +C_1(1-C_0^2)\lambda^3)\over 1+\lambda^6}\cr}\cr}$$
For the special case $C_0=0$, this simplifies, and is proportional to:
$${\bf 1}+\pmatrix{1& {-2C_1\lambda\over 1+\lambda^6}&0\cr
{-2C_1\lambda^5\over 1+\lambda^6}&1&0\cr
0&0&{2C_1\lambda^3\over 1+\lambda^6}\cr}.$$
The latter satisfies
$$\K^\dagger (\lambda )=\K( 1/\lambda ).$$

\appendix{D}{The other non simply-laced cases}

The implications of eq\ordertwoa\ for the non simply-laced root systems have
been
analysed and are listed in this appendix ($B_i=2C_i$). The corresponding
$\K$-matrices have
not been calculated for most of the cases. The simple solution \trivial\ is
always a
possibility and will not be listed separately for each case.
\bigskip
\noindent $a_{2n}^{(2)}$ ($n>2$):
$$\matrix{\hbox{\bf either\qquad}\ &C_i=\pm 1\quad&0\le i\le n-1, \quad &C_n\
\hbox{arbitrary},\cr
\hbox{\bf or\qquad}\ &C_i=0\quad &1\le i\le n,\quad &C_0\
\hbox{arbitrary},\cr}$$
where $C_n$ is the shortest simple root.
\bigskip
\noindent $a_{4}^{(2)}$:
$$\matrix{\hbox{\bf either\qquad}\ &C_0,C_1=\pm 1,&\quad&C_2\
\hbox{arbitrary},\cr
\hbox{\bf or\qquad}\ &C_0=\pm 1,\ &C_2=0,\quad &C_1\ \hbox{arbitrary},\cr
\hbox{\bf or\qquad}\ &C_1,C_2=0,&\quad& C_0\ \hbox{arbitrary},\cr}$$
\bigskip
\noindent $b_n^{(1)}$:
$$C_i=\pm 1 \qquad 0\le i\le n-1, \qquad C_n\ \hbox{arbitrary},$$
where $n$ labels the short simple root.
\bigskip
\noindent $a_{2n-1}^{(2)}$:
$$\matrix{\hbox{\bf either\qquad}\ &C_i=\pm 1\quad \hbox{for\ all}\ i,&\quad\cr
\hbox{\bf or\qquad}\ &C_i=0\quad 0\le i\le n-1,\quad&C_n\
\hbox{arbitrary},\cr}$$
where $n$ labels the long simple root.
\bigskip
\noindent $c_n^{(1)}$:
$$\matrix{\hbox{\bf either\qquad}\ &C_i=\pm 1\quad \hbox{for\ all}\
i,&\quad&\cr
\hbox{\bf or\qquad}\ &C_i=0\quad 1\le i\le n-1,&\quad &C_0,C_n\
\hbox{arbitrary},\cr}$$
where $n$ labels the long simple root.
\bigskip
\noindent $d_n^{(2)}$:
$$C_i=\pm 1\quad 1\le i\le n-1,\quad C_0,C_n\ \hbox{arbitrary},$$
where $n$ labels the short simple root.
\bigskip
\noindent $g_2^{(1)}$
$$C_0,C_1=\pm 1,\quad C_2\  \hbox{arbitrary},$$
where $2$ labels the short root.
\bigskip
\noindent $d_4^{(3)}$:
$$\matrix{\hbox{\bf either\qquad}\ &C_i=\pm 1\quad \hbox{for\ all}\
i,&\quad&\cr
\hbox{\bf or\qquad}\ &C_0,C_1=0,\quad &C_2\ \ \ \hbox{arbitrary}\cr},$$
where 2 labels the long simple root.
\bigskip
\noindent $f_4^{(1)}$
$$\matrix{\hbox{\bf either\qquad}\ &C_i=\pm 1\quad \hbox{for\ all}\
i,&\quad&\cr
\hbox{\bf or\qquad}\ &C_i=\pm 1\quad 0\le i\le 2,&\quad &C_3,C_4=0,\cr}$$
where $3,4$ label the short simple roots.
\bigskip
\noindent $e_6^{(2)}$:
$$\matrix{\hbox{\bf either\qquad}\ &C_i=\pm 1\quad \hbox{for\ all}\
i,&\quad&\cr
\hbox{\bf or\qquad}\ &C_i=0\quad 0\le i\le 2,&\quad &C_3,C_4=\pm 1,\cr}$$

\listrefs
\end